\documentclass[preprint,aps,eqsecnum]{revtex4} 
\usepackage{bm,amsfonts}

\begin{document}


\def\eqref#1{(\ref{#1})}
\def\eqrefs#1#2{(\ref{#1}) and~(\ref{#2})}
\def\eqsref#1#2{(\ref{#1}) to~(\ref{#2})}

\def\Eqref#1{Eq.~(\ref{#1})}
\def\Eqrefs#1#2{Eqs.~(\ref{#1}) and~(\ref{#2})}
\def\Eqsref#1#2{Eqs.~(\ref{#1}) to~(\ref{#2})}
\def\Eqpartref#1#2{Eq.~(\ref{#1}{#2})}

\def\secref#1{Sec.~\ref{#1}}
\def\secrefs#1#2{Sec.~\ref{#1} and~\ref{#2}}
\def\secsref#1#2{Sec.~\ref{#1} to~\ref{#2}}

\def\appref#1{App.~\ref{#1}}
\def\apprefs#1#2{App.~\ref{#1} and~\ref{#2}}
\def\appsref#1#2{App.~\ref{#1} to~\ref{#2}}

\def\Ref#1{Ref.~\cite{#1}}
\def\Refs#1{Refs.~\cite{#1}}

\def\Cite#1{${\mathstrut}^{\cite{#1}}$}

\def\tableref#1{Table~\ref{#1}}

\def\figref#1{Fig.~\ref{#1}}

\hyphenation{Eq Eqs Sec App Ref Fig}

\def\proclaim#1{\medbreak
\noindent{\it {#1}}}
\def\Proclaim#1#2{\medbreak
\noindent{\bf {#1}}{\it {#2}}\par\medbreak}

\def\EQ{\begin{equation}}
\def\EQs{\begin{eqnarray}}
\def\endEQ{\end{equation}}
\def\endEQs{\end{eqnarray}}

\def\proclaim#1{\medbreak
\noindent{\it {#1}}\par\medbreak}
\def\Proclaim#1#2{\medbreak
\noindent{\bf {#1}}{\it {#2}}\par\medbreak}

\def\Lemma{{\it Lemma\ }}
\def\Theorem#1{{\bf Theorem #1}}

\def\endproof{
\setbox2=\hbox{{$\sqcup$}} \setbox1=\hbox{{$\sqcap$}} 
\dimen1=\wd1
\box2\kern-\dimen1 \hbox to\dimen1{\box1} }

\def\newline{\hfil\break}
\def\fewquad{\qquad\qquad}
\def\severalquad{\qquad\fewquad}
\def\manyquad{\qquad\severalquad}
\def\manymanyquad{\manyquad\manyquad}


\def\ontop#1#2{
\setbox2=\hbox{{$#2$}} \setbox1=\hbox{{$\scriptscriptstyle #1$}} 
\dimen1=0.5\wd2 
\advance\dimen1 by 0.5\wd1 
\dimen2=1.4\ht2
\ifdim\wd1>\wd2 \raise\dimen2\box1 \kern-\dimen1 \hbox to\dimen1{\box2\hfill}
\else \box2\kern-\dimen1 \raise\dimen2 \hbox to\dimen1{\box1\hfill} \fi }

\def\onbot#1#2{
\setbox2=\hbox{{$#2$}} \setbox1=\hbox{{$\scriptscriptstyle #1$}} 
\dimen1=0.5\wd2 
\advance\dimen1 by 0.5\wd1 
\dimen2=1.2\ht1
\ifdim\wd1>\wd2 \lower\dimen2\box1 \kern-\dimen1 \hbox to\dimen1{\box2\hfill}
\else \box2\kern-\dimen1 \lower\dimen2 \hbox to\dimen1{\box1\hfill} \fi }

\def\eqtext#1{\hbox{\rm{#1}}}


\def\mixedindices#1#2{{\mathstrut}^{#1}_{#2}}
\def\downindex#1{{\mathstrut}^{\mathstrut}_{#1}}
\def\upindex#1{{\mathstrut}_{\mathstrut}^{#1}}
\def\downupindices#1#2{\downindex{#1}\upindex{#2}}
\def\updownindices#1#2{\upindex{#1}\downindex{#2}}
\def\sub#1{{}_{#1}}
\def\smallscript{\scriptscriptstyle}

\def\tensor#1#2#3{{#1}\mixedindices{#2}{#3}}
\def\covector#1#2{{#1}\downindex{#2}}
\def\vector#1#2{{#1}\upindex{#2}}

\def\id#1#2{\delta\downupindices{#1}{#2}}
\def\cross#1#2{\epsilon\,\updownindices{#1}{#2}}
\def\vol#1{\epsilon\,\downindex{#1}}
\def\invvol#1{\epsilon\,\upindex{#1}}
\def\cross#1#2{\epsilon\,\downupindices{#1}{#2}}

\def\flat#1{\eta\downindex{#1}}
\def\invflat#1{\eta\upindex{#1}}

\def\measure#1{d\,x^{#1}}

\def\frame#1#2#3{e\updownindices{#1}{#2}\upindex{#3}}
\def\invframe#1#2#3{{e^{\smallscript -1}}\updownindices{#2}{#1}\downindex{#3}}
\def\sodder#1#2{\sigma\updownindices{#1}{#2}}
\def\invsodder#1#2{\sigma\downupindices{#1}{#2}}

\def\vole#1#2#3{e\downupindices{#1}{#2}\downindex{#3}}
\def\dete#1#2{det(e)\updownindices{#1}{#2}}

\def\g#1#2{g\downupindices{#1}{#2}}
\def\invg#1#2{g^{-1}\updownindices{#1}{#2}}
\def\detg#1#2{det^{\frac{1}{2}}(g)\updownindices{#1}{#2}}

\def\conn#1#2#3{\Gamma\upindex{#1}\downupindices{#2}{#3}}
\def\R#1#2{R\downupindices{#1}{#2}}
\def\G#1#2{G\downupindices{#1}{#2}}
\def\T#1#2{T\downupindices{#1}{#2}}

\def\weyl#1#2#3{\varphi\mixedindices{#1}{#2}\upindex{#3}}
\def\cweyl#1#2#3{{\bar \varphi}\mixedindices{#1}{#2}\upindex{#3}}

\def\w#1#2{\omega\downupindices{#1}{#2}}
\def\F#1#2#3{F\mixedindices{#1}{#2}\upindex{#3}}
\def\cw#1#2{{\bar \omega}\downupindices{#1}{#2}}
\def\cF#1#2#3{{\bar F}\mixedindices{#1}{#2}\upindex{#3}}

\def\lagr#1#2{L\mixedindices{#1}{\rm #2}}
\def\S#1{S\upindex{#1}}

\def\der#1{\partial\downindex{#1}}
\def\coder#1{\partial\upindex{#1}}
\def\spinder#1#2{\partial\downupindices{#1}{#2}}
\def\extder#1{d^{}_{#1}}

\def\covder#1{\nabla^{ }_{#1}}
\def\covcoder#1{\nabla_{}^{#1}}

\def\D#1{D\downindex{#1}}
\def\coD#1{D\upindex{#1}}

\def\parder#1#2{{\partial{#2}/\partial {#1}}}
\def\Parder#1#2{{\partial{#2} \over\partial{#1}}}

\def\ELoph#1#2#3{{\rm E}_h\mixedindices{#2}{#1}\big(#3\big)}
\def\ELoprs#1#2#3{{\rm E}_\psi\mixedindices{#2}{#1}\big(#3\big)}

\def\Lie#1{{\cal L}_{#1}}
\def\LieE#1#2#3{(\Lie{\delta}E)_{#3}\mixedindices{#2}{#1}}


\def\h#1#2{h\downupindices{#1}{#2}}
\def\rs#1#2{\psi\downupindices{#1}{#2}}
\def\crs#1#2{\bar\psi\downupindices{#1}{#2}}
\def\barh#1#2{\gamma\downupindices{#1}{#2}}

\def\algh#1#2{{\bm h}\downupindices{#1}{#2}}
\def\algrs#1#2{{\bm \psi}\downupindices{#1}{#2}}
\def\algframe#1#2{{\bm e}\downupindices{#1}{#2}}
\def\algweyl#1#2{{\bm \varphi}\downupindices{#1}{#2}}
\def\algmet#1#2{{\bm g}\downupindices{#1}{#2}}

\def\vect#1#2{\xi\downupindices{#1}{#2}}
\def\spin#1#2{\zeta\downupindices{#1}{#2}}
\def\cspin#1#2{\bar\zeta\downupindices{#1}{#2}}

\def\lorspin#1#2{\chi\downupindices{#1}{#2}}
\def\clorspin#1#2{\bar\chi\downupindices{#1}{#2}}

\def\varh#1#2#3#4{\ontop{(#2)}{\delta_{#1}}\h{#3}{#4}}
\def\varrs#1#2#3#4{\ontop{(#2)}{\delta_{#1}}\rs{#3}{#4}}

\def\var#1#2{\ontop{(#1)}{\delta_{#2}}}
\def\varcomm#1#2#3{\ontop{(#1)}{[\delta_{#2},\delta_{#3}]}}
\def\diffvar#1#2{\ontop{(#1)}{\Delta\delta_{#2}}}
\def\varbot#1#2{{\onbot{#1}{\delta}}{}_{#2}}
\def\vartopbot#1#2#3{\ontop{(#1)}{\varbot{#2}{#3}}}
\def\commtop#1#2#3{\ontop{(#1)}{[{#2},{#3}]}}

\def\Etop#1#2#3#4{\ontop{(#1)}{E_{#4}}\mixedindices{#3}{#2}}
\def\cEtop#1#2#3#4{\ontop{(#1)}{{\bar E}_{#4}}\mixedindices{#3}{#2}}

\def\Ltop#1{\ontop{(#1)}{L}}

\def\E#1#2#3{\tensor{E_{#3}}{#2}{#1}}
\def\cE#1#2#3{\tensor{{\bar E}_{#3}}{#2}{#1}}


\def\a#1#2{a\updownindices{#1}{#2}}
\def\n#1#2{c\updownindices{#1}{#2}}
\def\l#1#2{b\updownindices{#1}{#2}}
\def\m#1#2{d\updownindices{#1}{#2}}
\def\cn#1#2{\bar c\updownindices{#1}{#2}}
\def\cl#1#2{\bar b\updownindices{#1}{#2}}
\def\cm#1#2{\bar d\updownindices{#1}{#2}}

\def\A#1#2{A\updownindices{#1}{#2}}
\def\N#1#2{C\updownindices{#1}{#2}}
\def\L#1#2{B\updownindices{#1}{#2}}
\def\cN#1#2{\bar C\updownindices{#1}{#2}}
\def\cL#1#2{\bar B\updownindices{#1}{#2}}

\def\q#1#2{q\downupindices{#1}{#2}}
\def\cq#1#2{\bar q\downupindices{#1}{#2}}
\def\Q#1#2{Q\downupindices{#1}{#2}}
\def\cQ#1#2{\bar Q\downupindices{#1}{#2}}
\def\othq#1#2{q'\downupindices{#1}{#2}}
\def\cothq#1#2{{\bar q}'\downupindices{#1}{#2}}
\def\othQ#1#2{Q'\downupindices{#1}{#2}}
\def\cothQ#1#2{{\bar Q}'\downupindices{#1}{#2}}

\def\dualq#1#2{q^*\downupindices{#1}{#2}}

\def\X{{\bf X}}
\def\Y{{\bf Y}}
\def\x#1{\covector{\bf x}{#1}}
\def\y#1{\covector{\bf y}{#1}}

\def\unit#1{\openone\upindex{#1}}

\def\ind#1{{\cal #1}}
\def\indsub#1#2{{\cal #1}_{#2}}

\def\sgA#1{\mathbb{A}_{\rm SG}^{\rm #1}}
\def\gA{\mathbb{A}_{\rm G}}
\def\multA#1{\mathbb{A}_{\rm internal}^{\rm #1}}
\def\hrsA#1{\mathbb{A}_{\rm coupling}^{\rm #1}}
\def\genA#1{\mathbb{A}^{\rm #1}}

\def\evenG#1{\mathbb{G}^{\rm even}_{\rm #1}}
\def\oddG#1{\mathbb{G}^{\rm odd}_{\rm #1}}
\def\gradedG#1{\mathbb{G}_{\rm #1}}
\def\V#1{\mathbb{V}_{\rm #1}}


\def\iso{\simeq}
\def\into{\rightarrow}

\def\span#1{\rm{span\ }\{{#1}\}}
\def\ad{^{\rm T}}
\def\hermad{^{\dagger}}
\def\inv{{}^{-1}}

\def\const{{\rm const.}}

\def\Rnum{{\mathbb R}}
\def\Cnum{{\mathbb C}}

\def\frac#1#2{{\textstyle {#1 \over #2}}}

\def\cc#1{
\setbox1=\hbox{${#1}$} 
\ifdim\wd1>9pt \overline{#1} \else \bar{#1} \fi }

\def\bigcc#1{\overline{\mathstrut #1}}


\def\grino/{spin-3/2}
\def\gron/{spin-2}
\def\Grino/{Spin-3/2}
\def\Gron/{Spin-2}

\def\EL/{Euler\discretionary{-}{}{-}Lagrange}

\def\Eeq/{Einstein's equation}
\def\RSeq/{Rarita-Schwinger equation}
\def\EC/{Einstein\discretionary{-}{}{-}Cartan}
\def\RS/{Rarita\discretionary{-}{}{-}Schwinger}
\def\FP/{Fierz\discretionary{-}{}{-}Pauli}
\def\KR/{Kalb\discretionary{-}{}{-}Ramond}
\def\Gra/{Grassmann}
\def\Maj/{Majorana}
\def\Mnk/{Minkowski}
\def\susy/{supersymmetry}
\def\Nsusy/{N$=1$ supersymmetry}

\def\NSG/{N$=1$ supergravity}
\def\SG/{supergravity}
\def\GR/{general relativity}
\def\YM/{Yang\discretionary{-}{}{-}Mills}

\def\algv/{algebra\discretionary{-}{}{-}valued}

\def\ie/{$\rm i.e.$}
\def\eg/{$\rm e.g.$}
\def\etc/{$\rm etc.$}

\hyphenation{
all
along
anti
anti-com-mu-ta-tiv-i-ty
ap-pen-dix 
com-mu-ta-tiv-i-ty
con-straint
co-tan-gent
equa-tion
equa-tion-s
equiv-a-lent
evo-lu-tion
fields
form
iden-ti-ty
iden-ti-ties
im-por-tant
its
La-grang-ian
La-grang-ian-s
next
nev-er
prod-uct
real
sca-lar
shall
smooth
space-s
space-time
space-times
steps
strict
su-per-grav-i-ty
su-per-sym-me-try
sym-me-try
sym-me-tries
tak-en
tan-gent
term
two
use
use-s
vari-a-tion
vari-a-tion-s
}

\hyphenation{
Grass-ma-nn
Ein-stein
Min-kow-ski
}


\title{ On multi-graviton and multi-gravitino gauge theories }

\author{ Stephen C. Anco }
\affiliation{
Department of Mathematics, Brock University, St. Catharines, ON Canada L2S 3A1}
\email{sanco@brocku.ca}
\date{\today}

\begin{abstract}
This paper studies nonlinear deformations of 
the linear gauge theory of any number of \gron/ and \grino/ fields
with general formal multiplication rules 
in place of standard \Gra/ rules for manipulating the fields,
in four spacetime dimensions. 
General possibilities for multiplication rules and coupling constants
are simultaneously accommodated by regarding the set of fields
equivalently as a single algebra-valued \gron/ field and
single algebra-valued \grino/ field,
where the underlying algebra is factorized into 
a field-coupling part and an internal multiplication part. 
The condition that there exist a gauge invariant Lagrangian 
(to within a divergence) 
for these algebra-valued fields
is used to derive determining equations 
whose solutions give all allowed deformation terms, 
yielding nonlinear field equations and nonabelian gauge symmetries,
together with all allowed formal multiplication rules 
as needed in the Lagrangian
for demonstration of invariance under the gauge symmetries
and for derivation of the field equations. 
In the case of \gron/ fields alone, 
the main result of this analysis is that all deformations
(without any higher derivatives than appear in the linear theory)
are equivalent to an algebra-valued Einstein gravity theory. 
By a systematic examination of factorizations of the algebra, 
a novel type of nonlinear gauge theory of two or more \gron/ fields is found,
where the coupling for the fields is based on structure constants of
an anticommutative, anti-associative algebra, 
and with formal multiplication rules that make the fields anticommuting
(while products obey anti-associativity). 
Supersymmetric extensions of these results are obtained 
in the more general case when \grino/ fields are included. 
\end{abstract}

\maketitle
\newpage

\section{ Introduction and summary }

It has long been known \cite{gravity1,gravity2,gravity3,gravity4}
that the Einstein gravitational field equations
describe a nonlinear gauge theory of a massless \gron/ field 
(\ie/ graviton) 
as defined by the difference of the gravitational metric tensor
and any fixed background flat metric, 
where the gauge symmetry arises from diffeomorphisms 
on the metric field tensor. 
Moreover, the uniqueness of this theory in providing a consistent 
nonlinear self-coupling for a \gron/ field 
is by now well established from many points of view
\cite{uniqueness1,uniqueness2,uniqueness3,uniqueness4,uniqueness5,
uniqueness6,uniqueness7}. 
Nevertheless, there has been some interest in recent years 
in the possibility of consistent novel 
nonlinear gauge theories of \gron/ fields. 
This paper significantly elaborates one avenue of work on such possibilities.

The motivation is explained by certain features of classical \SG/ theories
\cite{sg1,sg2,sg3}, 
viewed as a supersymmetric extension of the Einstein gravity theory
involving, as a matter source, 
a massless \grino/ field (\ie/ gravitino) 
that is required to be formally anticommuting. 
In this extension, 
the massless \gron/ field remains formally commuting,
while products of the \gron/ and \grino/ fields are manipulated by
formal rules of \Gra/ multiplication. 
These rules, which serve as classical counterparts of graviton-gravitino
quantum field commutation relations, 
are used for manipulations in the Lagrangian 
to establish \susy/ invariance and to derive the field equations. 
\Nsusy/ involves a single pair of \gron/ and \grino/ fields, 
but for N$>1$ a complete \susy/ multiplet \cite{sg3} requires more fields,
and notably, additional \gron/ fields are needed if N$>8$. 

These features naturally suggest exploring the possibilities 
for nonlinear gauge theories of, firstly, 
a set of at least two ordinary (commuting) \gron/ fields,
and secondly, 
a single \gron/ field with formal multiplication rules
that make it noncommuting. 
Mathematically, note that 
a set of $n\ge 1$ ordinary \gron/ fields is equivalent to 
a single \gron/ field possessing an internal structure of 
a real $n$-dimensional vector space. 
Consequently, in a nonlinear theory
the coupling constants that appear in the Lagrangian 
for a set of ordinary \gron/ fields
thereby correspond to an algebraic structure on the internal vector space
possessed by an equivalent single \algv/ \gron/ field.  
The resulting algebraic structure, furthermore, 
serves to provide multiplication rules for manipulating products of
the \algv/ \gron/ field in the Lagrangian. 
This discussion shows that there is a well-defined mathematical equivalence
between nonlinear theories of 
a set of ordinary (\ie/ real-valued) \gron/ fields
and nonlinear theories of a single \gron/ field 
with formal multiplication rules 
represented by an internal algebra in which the \gron/ field takes values. 
In particular, under this point of view,
Einstein gravity theory for a single commuting \gron/ field
formulated using even-\Gra/ multiplication rules is the same as
a nonlinear theory found by Cutler~and Wald \cite{Cutler-Wald,Wald}
for a set of real-valued \gron/ fields 
with coupling constants corresponding to any even-\Gra/ algebra. 
More precisely, the equivalence here is such that
each \gron/ field is associated with a basis element in this algebra
\cite{Wald}. 

These considerations carry over in an obvious way to 
\grino/ (and other) fields. 
As a result, \NSG/ theory for 
a single commuting \gron/ field and a single anticommuting \grino/ field
based on \Gra/ multiplication rules
has an equivalent formulation as a nonlinear theory of 
a set of ordinary (real-valued) \gron/ and \grino/ fields
whose coupling constants are associated with any \Gra/ algebra
\cite{Isenberg1,Annalspaper}.
This motivates a fundamental question of whether classical \NSG/ theory
formulated in terms of a \Gra/ algebra 
is the unique possibility for a nonlinear gauge theory of 
a set of ordinary \gron/ and \grino/ fields;
and likewise, whether its graviton sector given by the formulation of
Einstein gravity theory involving an even-\Gra/ algebra 
is the unique possibility for a nonlinear gauge theory of 
a set of ordinary \gron/ fields alone. 

The most systematic approach for determining all possibilities
for nonlinear \gron//\grino/ gauge theories is by 
a deformation analysis of linear abelian gauge theory 
of a set of ordinary \gron/ and \grino/ fields. 
Here, deformations refer to 
adding quadratic and higher power terms in the linear field equations
while adding linear and higher power terms in the abelian gauge symmetries,
such that there exists a gauge invariant Lagrangian 
(to within a total divergence),
with the undeformed Lagrangian of the linear abelian theory 
not being equivalent to the deformed Lagrangian 
under field redefinitions. 
The condition of gauge invariance can be used to obtain 
determining equations to solve for the allowed form of the deformation terms
order by order in powers of the fields
(see \Refs{AMSpaper,Henneaux1}).
Two natural restrictions on the general form considered for these terms
come from requiring that the deformations preserve 
the number of gauge degrees of freedom
and initial-data degrees of freedom for the fields. 
This leads to 
restricting any derivatives 
in the deformed gauge symmetries and field equations 
to be of no higher order than those in the linear theory \cite{Annalspaper},
which will then be referred to as a non-higher-derivative deformation. 

A complete analysis of non-higher-derivative deformations 
for a set of arbitrarily many \gron/ and \grino/ fields
in four spacetime dimensions, 
without further restrictions or special assumptions on possible forms
for gauge symmetries and field equations, 
was first carried out in \Ref{Annalspaper}
using a field theoretic formulation of the deformation determining equations.
This analysis obtained two strong uniqueness results:
First, for a set of \gron/ fields alone,
the only non-higher-derivative deformations are equivalent to 
Einstein gravity theory for an even-\Gra/ \algv/ \gron/ field. 
However, for a set of \gron/ and \grino/ fields,
the allowed non-higher-derivative deformations correspond to 
a chiral generalization of classical \NSG/ theory
for an \algv/ pair of \gron/ and \grino/ fields
involving a novel modification of a \Gra/ algebra
such that the \grino/ field is only anticommutative
in combination with charge conjugation,
while the \gron/ field remains commutative, 
but commutativity (and associativity) of the \gron/ field 
in products with the \grino/ field 
holds only in combination with charge conjugation. 
The origin of this generalization stems from 
the Weyl-spinor formulation of \SG/ theory, 
in which the anticommuting nature of the \grino/ field 
is found to be never needed separately from charge conjugation
\cite{PhysRevpaper}.
This allows dispensing with certain \Gra/ multiplication rules,
and thereby defines a non-\Gra/ multiplication 
intertwined with charge conjugation. 
As a consequence, the \SG/ field equations and \susy/ 
become deformed by some chiral terms that otherwise would vanish
if the \grino/ field were strictly anticommuting. 
These terms essentially maintain the gauge invariance
for the deformation. 
The formulation of the resulting chiral generalized \NSG/ Lagrangian 
\cite{PhysRevpaper}
in terms of a pair of non-\Gra/ \algv/ \gron/ and \grino/ fields
is explained in \Ref{Annalspaper}.

It is worth emphasizing that, due to gauge invariance, 
the coupling of the \grino/ field to the \gron/ field
in this generalized \NSG/ theory is completely consistent.
Moreover, when viewed as a nonlinear gauge theory of 
a pair of \gron/ and \grino/ fields with formal internal multiplication rules,
it shares the same key features as standard classical \NSG/ theory ---
well-posedness of the initial value problem \cite{Isenberg2},
formal positive energy properties \cite{energy},
and a geometrical description \cite{sg4}
in terms of curvature of the metric tensor associated to the \gron/ field,
with a matter source and torsion determined by the \grino/ field. 
In the corresponding formulation 
for a set of ordinary \gron/ and \grino/ fields,
the field equations continue to be well-posed
and retain a geometrical meaning in terms of 
an \algv/ metric and curvature given by the \gron/ fields,
as discussed in detail in \Ref{Wald},
and an \algv/ torsion and matter source 
determined by the \grino/ fields,
outlined in \Ref{Annalspaper}.
However, the field equations viewed in this manner
have a partially decoupled nonlinear structure,
where the coupling terms reflect the algebra multiplication relations
among the basis elements of the internal algebra \cite{Isenberg1}. 
As a result of this structure, 
the canonical stress-energy tensor obtained from the Lagrangian
for the set of field equations 
is found to yield a total energy that, in general, is of indefinite sign. 
(This feature has been studied in \Ref{Anco-Wald}
for an analogous nonlinear theory of a set of scalar fields,
describing a standard quartic self-coupling 
for an equivalent \algv/ single scalar field,
with a simple choice of internal algebra.)

The non-positivity of energy for these coupled 
ordinary \gron/ and \grino/ fields
is directly related to the multiplication being nontrivial 
in the internal algebra on which the nonlinear gauge theory is based. 
Indeed, in \Ref{Henneaux2}
it was subsequently shown that in the case of a set of \gron/ fields
the only internal algebra yielding a nonlinear gauge theory
whose canonical energy is positive 
is given by a direct sum of one-dimensional unit algebras. 
(This feature extends to the more general case of 
a set of \gron/ and \grino/ fields \cite{thesis}.)
\Ref{Henneaux2} also gave a deformation analysis that 
generalized the uniqueness result 
for even-\Gra/ valued Einstein gravity theory
as a nonlinear gauge theory for a set of ordinary \gron/ fields,
by relaxing the natural restriction on highest order derivatives 
in deformations of the \gron/ gauge symmetries
(and allowing other than four spacetime dimensions)
through the use of powerful BRST cohomology techniques \cite{brst}
to formulate and solve the deformation determining equations. 

To-date, all previous investigations of nonlinear gauge theories
for a set of \gron/ fields, 
and more generally, 
a set of \gron/ and \grino/ fields, 
have considered only ordinary (\ie/ real-valued) fields. 
This excludes, consequently, 
the possibility of more than one anticommuting \grino/ field
with formal odd-\Gra/ multiplication rules,
as would arise in N$>1$ \susy/ multiplets. 
It also excludes the more exotic possibility of a noncommuting \gron/ field. 
The main purpose of the present paper is to fill the previous gaps
by giving a systematic determination of 
all possible nonlinear gauge theories of an arbitrary number of 
\gron/ and \grino/ fields each with its own 
internal formal multiplication rules,
and as a special case, 
\gron/ fields with other than even-\Gra/ multiplication rules,
in four spacetime dimensions. 
The results yield an exotic classical nonlinear gauge theory of
anticommuting \gron/ fields,
and a supersymmetric extension with commuting \grino/ fields, 
employing formal multiplication rules 
in which the usual spin-statistics relation for the \gron//\grino/ fields 
is reversed at the classical level. 
(These nonlinear theories turn out to involve 
only products of distinct fields,
but never a product of any field with itself,
so anticommutativity is not needed to hold for each individual \gron/ field
and likewise for commutativity of each individual \grino/ field,
in remarkable accordance with 
the general spin-statistics relations \cite{spin} 
allowed by quantum field theory.)
In \secref{deformations}
the deformation analysis used to find these theories is summarized,
and uniqueness results from this analysis are stated. 
The theories are presented in detail in \secref{results}. 
Some features of the theories are discussed 
along with a few concluding remarks in \secref{remarks}.
An appendix summarizes some material on relevant algebras.

\section{ Deformation analysis }
\label{deformations}

We begin by setting up the framework for deformations, 
using a generalization of the formalism (and notation) of \Ref{Annalspaper}
to accommodate any number of \gron/ and \grino/ fields
with an internal vector space structure 
for any formal multiplication rules.

\subsection{ Preliminaries }

The \gron/ fields are taken to be real spinorial tensors 
$\h{aBB'}{\mu}$, $\mu=1,\ldots,n$,
and the \grino/ fields are taken to be complex vector-spinors 
$\rs{aB}{\Lambda}$, $\Lambda=1,\ldots,n'$.
This choice of variables is motivated by 
the chiral spinorial formulation of classical \NSG/ theory 
\cite{variables} 
that uses a null spinorial tetrad $\frame{BB'}{a}{}$
and a Weyl vector-spinor $\weyl{B}{a}{}$
which are, respectively, even and odd \Gra/-valued. 
Linearization about a flat tetrad $\invsodder{aBB'}{}$
and a zero vector-spinor 
in that theory yields 
a \gron/ field $\h{aBB'}{}=\frame{}{aBB'}{}- \invsodder{aBB'}{}$
and a \grino/ field $\rs{aB}{}=\weyl{}{aB}{}$,
where $n=n'=1$. 

Each field $\h{aBB'}{\mu},\rs{aB}{\Lambda}$ here is regarded as
taking values in an internal vector space $\X,\Y$,
where we fix $\X$ to be a real vector space of some arbitrary dimension
with a basis $\x{1},\x{2},\ldots$,
and $\Y$ to be a complexified vector space of some arbitrary dimension
with a basis $\y{1},\y{2},\ldots$, 
respectively. 
The structure necessary to formulate internal multiplication rules
for $\h{aBB'}{\mu}$ and $\rs{aB}{\Lambda}$
will be given by multilinear maps from products of $\X,\Y$, 
into $\X$ and $\Y$.
Note that the coefficients in a basis expansion of such products
define multiplication structure constants, 
which represent the multiplication rules. 
With respect to these bases, 
we expand the fields
\EQ
\h{aBB'}{\mu} 
= \h{aBB'}{\mu,1} \x{1} +\h{aBB'}{\mu,2} \x{2} + \cdots ,\quad
\rs{aB}{\Lambda} 
=\rs{aB}{\Lambda,1} \y{1} +  \rs{aB}{\Lambda,2} \y{2} +\cdots
\label{fieldset}
\endEQ
where $\h{aBB'}{\mu,1},\h{aBB'}{\mu,2},\ldots$,
$\rs{aB}{\Lambda,1},\rs{aB}{\Lambda,2},\ldots$
are ordinary real-valued spinorial tensor fields
and complex-valued vector-spinor fields.
Products of $\h{aBB'}{\mu}$ and $\rs{aB}{\Lambda}$ 
involving internal multiplication rules then reduce to 
ordinary products of 
$\h{aBB'}{\mu,1},\h{aBB'}{\mu,2},\ldots$
and $\rs{aB}{\Lambda,1},\rs{aB}{\Lambda,2},\ldots$
as specified by the multiplication structure constants. 
This allows the set of ordinary \gron/ and and \grino/ fields
$\h{aBB'}{\mu,\ind{A}},\rs{aB}{\Lambda,\ind{A'}}$ 
($\ind{A}=1,2,\ldots, \ind{A'}=1,2,\ldots$)
to be used as the field variables 
for the subsequent framework here. 
For this purpose it is convenient hereafter 
to employ the multi-index notation 
$\indsub{A}{\mu}=(\mu,\ind{A})$, $\indsub{A}{\Lambda}'=(\Lambda,\ind{A'})$. 
The summation convention with respect to a repeated index $\indsub{A}{\mu}$
will mean a sum over both $\mu$ and $\ind{A}$,
and similarly, 
a sum over both $\Lambda$ and $\ind{A'}$ 
for a repeated index $\indsub{A}{\Lambda}'$.

To proceed, we start from the linear abelian gauge theory 
for the set of $n\ge 1$ \gron/ fields and $n'\ge 1$ \grino/ fields 
each with an internal vector space structure \eqref{fieldset}, 
on a flat 4-dimensional spacetime manifold.
In terms of the ordinary field variables 
$\h{aBB'}{\indsub{A}{\mu}},\rs{aB}{\indsub{A}{\Lambda}'}$, 
the linear \gron/ field equations are given by 
the \FP/ equation 
\EQ
0= \coder{c}\der{c}\barh{ab}{\indsub{A}{\mu}} 
- 2\coder{c}\der{(a}\barh{b)c}{\indsub{A}{\mu}} 
+\flat{ab} \coder{c}\coder{d}\barh{cd}{\indsub{A}{\mu}}
= 4 \Etop{1}{ab}{\indsub{A}{\mu}}{h}
\label{linheq}
\endEQ
with 
$\barh{ab}{\indsub{A}{\mu}}= 
\h{ab}{\indsub{A}{\mu}} - \frac{1}{2}\flat{ab}\h{c}{c\indsub{A}{\mu}}$
where 
$\h{ab}{\indsub{A}{\mu}} = \invsodder{(b}{BB'}\h{a)BB'}{\indsub{A}{\mu}}$, 
and the linear \grino/ field equations are given by 
the \RS/ equation 
\EQ 
0= \cross{a}{bcd} \invsodder{bB}{B'} \der{c} \crs{dB'}{\indsub{A}{\Lambda}'}
= 2i \Etop{1}{aB}{\indsub{A}{\Lambda}'}{\psi}
\label{linrseq}
\endEQ
where $\invsodder{bBB'}{}$ is any flat spinorial tetrad
and $\vol{abcd} = 
2i \invsodder{[a|A'|}{A} \invsodder{b|B|}{A'} 
\invsodder{c|B'|}{B} \invsodder{d]A}{B'}$
is the associated volume tensor of 
the flat metric $\flat{ab}=\invsodder{a}{CC'} \invsodder{bCC'}{}$.
The abelian gauge symmetries on these field variables
consist of infinitesimal variations given by 
a linearized general-covariance symmetry 
\EQ
\varh{\xi}{0}{aBB'}{\indsub{A}{\mu}} 
= 2\sodder{b}{BB'} \der{(a}\vect{b)}{\indsub{A}{\mu}} ,\quad
\varrs{\xi}{0}{aB}{\indsub{A}{\Lambda}'} =0 ,
\label{lindiffeosymm}
\endEQ
a linearized \susy/ 
\EQ
\varrs{\zeta}{0}{aB}{\indsub{A}{\Lambda}'}
= \der{a}\spin{B}{\indsub{A}{\Lambda}'} ,\quad
\varh{\zeta}{0}{aBB'}{\indsub{A}{\mu}}  = 0 ,
\label{linsusysymm}
\endEQ
and a linearized local Lorentz symmetry
\EQ
\varh{\chi}{0}{aBB'}{\indsub{A}{\mu}} 
=\invsodder{aB'}{A} \lorspin{AB}{\indsub{A}{\mu}} + c.c. ,\quad
\varrs{\zeta}{0}{aB}{\indsub{A}{\Lambda}'} =0 ,
\label{linlorentzsymm}
\endEQ
which involve as respective variables
the arbitrary covector fields $\vect{a}{\indsub{A}{\mu}}$, 
spinor fields $\spin{B}{\indsub{A}{\Lambda}'}$,
and symmetric spinor fields $\lorspin{AB}{\indsub{A}{\mu}}
= \lorspin{(AB)}{\indsub{A}{\mu}}$.
The field equations have a gauge invariant Lagrangian formulation 
given by 
\EQ
\Ltop{2} = 
\frac{1}{2} \q{\indsub{A}{\mu}\indsub{B}{\nu}}{} 
\h{aBB'}{\indsub{A}{\mu}} \Etop{1}{}{aBB'\indsub{B}{\nu}}{h} 
+ \frac{1}{2}( \othq{\indsub{A}{\Lambda}'\indsub{B}{\Gamma}'}{}
\crs{aB'}{\indsub{A}{\Lambda}'} \cEtop{1}{}{aB'\indsub{B}{\Gamma}'}{\psi} 
+ c.c. )
\label{quadL}
\endEQ
where $\q{\indsub{A}{\mu}\indsub{B}{\nu}}{}$ 
and $\othq{\indsub{A}{\Lambda}'\indsub{B}{\Gamma}'}{}$
are, respectively, 
components of any fixed diagonal real-symmetric and skew-hermitian 
nondegenerate matrices. 
In particular, 
through \EL/ operators 
$\ELoph{aBB'}{\indsub{A}{\mu}}{\cdot}$ 
and $\ELoprs{aB}{\indsub{A}{\Lambda}'}{\cdot}$,
which annihilate total divergences, 
the Lagrangian \eqref{quadL} yields the field equations
\EQ
\Etop{1}{ab}{\indsub{A}{\mu}}{h} 
= \ELoph{ab}{\indsub{A}{\mu}}{ \Ltop{2} } ,\quad
\Etop{1}{aB}{\indsub{A}{\Lambda}'}{\psi}
= \ELoprs{aB}{\indsub{A}{\Lambda}'}{ \Ltop{2} } ,
\endEQ
while invariance with respect to the gauge symmetries is expressed by
\EQs
&&
\ELoph{aBB'}{\indsub{A}{\mu}}{ \var{0}{\xi}\Ltop{2} }
= \ELoph{aBB'}{\indsub{A}{\mu}}{ \var{0}{\zeta}\Ltop{2} }
= \ELoph{aBB'}{\indsub{A}{\mu}}{ \var{0}{\chi}\Ltop{2} }
=0 ,
\\
&&
\ELoprs{aB}{\indsub{A}{\Lambda}'}{ \var{0}{\xi}\Ltop{2} } 
= \ELoprs{aB}{\indsub{A}{\Lambda}'}{ \var{0}{\zeta}\Ltop{2} } 
= \ELoprs{aB}{\indsub{A}{\Lambda}'}{ \var{0}{\chi}\Ltop{2} } 
=0 .
\endEQs

A deformation of this linear theory is defined by 
adding terms of linear and higher powers 
to the abelian gauge symmetries
\EQ
\varh{}{0}{aBB'}{\indsub{A}{\mu}} 
+ \varh{}{1}{aBB'}{\indsub{A}{\mu}} +\cdots =
\delta_{} \h{aBB'}{\indsub{A}{\mu}} ,\quad
\varrs{}{0}{aB}{\indsub{A}{\Lambda}'} 
+ \varrs{}{1}{aB}{\indsub{A}{\Lambda}'} +\cdots =
\delta_{} \rs{aB}{\indsub{A}{\Lambda}'} ,
\label{deformsymms}
\endEQ
while simultaneously adding terms of quadratic and higher powers 
to the linear field equations
\EQs
\Etop{1}{aBB'}{\indsub{A}{\mu}}{h} 
+ \Etop{2}{aBB'}{\indsub{A}{\mu}}{h} + \cdots 
= \E{aBB'}{\indsub{A}{\mu}}{h} ,\quad
\Etop{1}{aB}{\indsub{A}{\Lambda}'}{\psi} 
+\Etop{2}{aB}{\indsub{A}{\Lambda}'}{\psi} + \cdots 
= \E{aB}{\indsub{A}{\Lambda}'}{\psi} ,
\label{deformeqs}
\endEQs
such that there exists a gauge invariant Lagrangian 
(to within a total divergence)
\EQ
\Ltop{2} +\Ltop{3} +\cdots = L
\label{deformL}
\endEQ
satisfying
\EQ
\ELoph{aBB'}{\indsub{A}{\mu}}{\delta_{} L } = 0 ,\quad
\ELoprs{aB}{\indsub{A}{\Lambda}'}{ \delta_{} L } = 0 ,
\label{gaugeinv}
\endEQ
where the Lagrangian is related to the field equations through 
$\ELoph{aBB'}{\indsub{A}{\mu}}{ L } = \E{aBB'}{\indsub{A}{\mu}}{h}$
and 
$\ELoprs{aB}{\indsub{A}{\Lambda}'}{ L } = \E{aB}{\indsub{A}{\Lambda}'}{\psi}$.
We restrict attention hereafter to non-higher-derivative deformations
whose terms are locally constructed from 
$\h{aBB'}{\indsub{A}{\mu}},\rs{aB}{\indsub{A}{\Lambda}'}$, 
and their derivatives, 
(in addition to the spacetime coordinates) 
such that at most one derivative in total of 
$\h{aBB'}{\indsub{A}{\mu}}$, $\rs{aB}{\indsub{A}{\Lambda}'}$, 
$\vect{a}{\indsub{A}{\mu}}$, 
$\spin{B}{\indsub{A}{\Lambda}'}$,
$\lorspin{AB}{\indsub{A}{\mu}}$ 
appears in the deformed gauge symmetries,
and at most two derivatives in total of 
$\h{aBB'}{\indsub{A}{\mu}},\rs{aB}{\indsub{A}{\Lambda}'}$
appear in the deformed field equations. 
Moreover, 
any such deformations that are related by 
invertible, nonlinear locally constructed field redefinitions
(\ie/ change of variables) 
of $\h{aBB'}{\indsub{A}{\mu}}$, $\rs{aB}{\indsub{A}{\Lambda}'}$, 
$\vect{a}{\indsub{A}{\mu}}$, $\spin{B}{\indsub{A}{\Lambda}'}$,
$\lorspin{AB}{\indsub{A}{\mu}}$ 
are considered to be equivalent. 
The condition of local gauge invariance \eqref{gaugeinv}
provides the determining equations for allowed deformations. 

The determining equations have a more useful and geometrical formulation
as the following Lie derivative equations. 
We introduce the Lie derivative operator $\Lie{\delta}$
with respect to field variations 
$(\delta\h{aBB'}{\indsub{A}{\mu}},\delta\rs{aB}{\indsub{A}{\Lambda}'})$
acting on field equations 
$(\E{aBB'}{\indsub{A}{\mu}}{h},\E{aB}{\indsub{A}{\Lambda}'}{\psi})$
by
\EQs
\LieE{\indsub{A}{\mu}}{aBB'}{h} = &&
\delta\E{\indsub{A}{\mu}}{aBB'}{h}
+ \E{\indsub{B}{\nu}}{cDD'}{h} 
\parder{\h{aBB'}{\indsub{A}{\mu}}}{ \delta\h{cDD'}{\indsub{B}{\nu}} }
+ \E{\indsub{B}{\Gamma}'}{cD}{\psi}
\parder{\h{aBB'}{\indsub{A}{\mu}}}{ \delta\rs{cD}{\indsub{B}{\Gamma}'} }
+ c.c.
\nonumber\\&&
-\der{e}\Big( \E{\indsub{B}{\nu}}{cDD'}{h} 
\parder{(\der{e}\h{aBB'}{\indsub{A}{\mu}})}
{ \delta\h{cDD'}{\indsub{B}{\nu}} }
+ \E{\indsub{B}{\Gamma}'}{cD}{\psi}
\parder{(\der{e}\h{aBB'}{\indsub{A}{\mu}})}
{ \delta\rs{cD}{\indsub{B}{\Gamma}'} } + c.c. \Big)
\nonumber\\&&\\
\LieE{\indsub{A}{\Lambda}'}{aB}{\psi} = &&
\delta\E{\indsub{A}{\Lambda}'}{aB}{\psi} 
+ \E{\indsub{B}{\Gamma}'}{cD}{\psi}
\parder{\rs{aB}{\indsub{A}{\Lambda}'}}{ \delta\rs{cD}{\indsub{B}{\Gamma}'} }
+ \cE{\indsub{B}{\Gamma}'}{cD'}{\psi}
\parder{\rs{aB}{\indsub{A}{\Lambda}'}}{ \delta\crs{cD'}{\indsub{B}{\Gamma}'} }
\nonumber\\&&
+ \E{\indsub{B}{\nu}}{cDD'}{h} 
\parder{\rs{aB}{\indsub{A}{\Lambda}'}}{ \delta\h{cDD'}{\indsub{B}{\nu}} }
-\der{e}\Big( \E{\indsub{B}{\nu}}{cDD'}{h} 
\parder{(\der{e}\rs{aB}{\indsub{A}{\Lambda}'})}
{ \delta\h{cDD'}{\indsub{B}{\nu}} }
\nonumber\\&&
+ \E{\indsub{B}{\Gamma}'}{cD}{\psi}
\parder{(\der{e}\rs{aB}{\indsub{A}{\Lambda}'})}
{ \delta\rs{cD}{\indsub{B}{\Gamma}'} } 
+ \cE{\indsub{B}{\Gamma}'}{cD'}{\psi}
\parder{(\der{e}\rs{aB}{\indsub{A}{\Lambda}'})}
{ \delta\crs{cD'}{\indsub{B}{\Gamma}'} } \Big)
\endEQs
where
\EQs
\delta = &&
\delta\h{cDD'}{\indsub{B}{\nu}} 
\parder{\h{cDD'}{\indsub{B}{\nu}}}{}
+ \delta\rs{cD}{\indsub{B}{\Gamma}'}
\parder{\rs{cD}{\indsub{B}{\Gamma}'}}{} 
+ c.c.
\nonumber\\&&
+ (\der{e}\delta\h{cDD'}{\indsub{B}{\nu}})
\parder{(\der{e}\h{cDD'}{\indsub{B}{\nu}})}{} 
+ (\der{e}\delta\rs{cD}{\indsub{B}{\Gamma}'})
\parder{(\der{e}\rs{cD}{\indsub{B}{\Gamma}'})}{}
+ c.c.
\endEQs
defines the field variation operator. 
Here, note, we have taken into account
the restrictions on highest orders of derivatives 
in the field equations and field variations
as relevant for non-higher-derivative deformations. 

\Proclaim{ Proposition~1. }{
Local gauge invariance holds iff 
the Lie derivative of the field equations
with respect to the gauge symmetries vanishes:
\EQ
\Lie{\delta_\xi}
(\E{aBB'}{\indsub{A}{\mu}}{h},\E{aB}{\indsub{A}{\Lambda}'}{\psi})
=0 ,\quad
\Lie{\delta_\zeta}
(\E{aBB'}{\indsub{A}{\mu}}{h},\E{aB}{\indsub{A}{\Lambda}'}{\psi})
=0 ,\quad
\Lie{\delta_\chi}
(\E{aBB'}{\indsub{A}{\mu}}{h},\E{aB}{\indsub{A}{\Lambda}'}{\psi})
=0 .
\label{liedereq}
\endEQ
}

These invariance equations assert geometrically that
the gauge symmetries are tangent directions to the surface
defined by solutions of the field equations
in the space of \gron/ and \grino/ field configurations. 
Gauge invariance implies, consequently, that
the commutations of the gauge symmetries 
have the same property. 

\Proclaim{ Proposition~2. }{
Local gauge invariance holds only if 
the Lie derivative of the field equations
with respect to the gauge symmetry commutators vanishes:
\EQs
&&
\Lie{[\delta_{\xi_1},\delta_{\xi_2}]}
(\E{aBB'}{\indsub{A}{\mu}}{h},\E{aB}{\indsub{A}{\Lambda}'}{\psi})
=0 ,\quad
\Lie{[\delta_{\zeta_1},\delta_{\zeta_2}]}
(\E{aBB'}{\indsub{A}{\mu}}{h},\E{aB}{\indsub{A}{\Lambda}'}{\psi})
=0 ,\quad
\Lie{[\delta_{\chi_1},\delta_{\chi_2}]}
(\E{aBB'}{\indsub{A}{\mu}}{h},\E{aB}{\indsub{A}{\Lambda}'}{\psi})
=0 ,
\nonumber\\
&&
\Lie{[\delta_{\xi_1},\delta_{\zeta_2}]}
(\E{aBB'}{\indsub{A}{\mu}}{h},\E{aB}{\indsub{A}{\Lambda}'}{\psi})
=0 ,\quad
\Lie{[\delta_{\xi_1},\delta_{\chi_2}]}
(\E{aBB'}{\indsub{A}{\mu}}{h},\E{aB}{\indsub{A}{\Lambda}'}{\psi})
=0 ,\quad
\Lie{[\delta_{\zeta_1},\delta_{\chi_2}]}
(\E{aBB'}{\indsub{A}{\mu}}{h},\E{aB}{\indsub{A}{\Lambda}'}{\psi})
=0 .
\nonumber\\
\label{liedercommeq}
\endEQs
}

An expansion of these equations \eqrefs{liedereq}{liedercommeq}
in powers of the fields
gives a hierarchy of determining equations whose solutions 
yield all allowed deformation terms 
in the field equations and gauge symmetries. 

Compared with \Ref{Annalspaper,Henneaux2},
this framework for deformations of
the linear abelian gauge theory of a set of 
ordinary \gron/ and \grino/ fields 
is more general in that it does not use the familiar choice of
a symmetric tensor for the \gron/ field variables,
corresponding to 
$\invsodder{[b}{BB'}\h{a]BB'}{\indsub{A}{\mu}} =0$, 
as imposed by gauge fixing with the linearized local Lorentz symmetry. 
Indeed, the choice here of nonsymmetric \gron/ field variables
$\h{aBB'}{\indsub{A}{\mu}}$ 
allows the framework to encompass the related possibilities of
deforming the local Lorentz symmetry on the fields 
$\h{aBB'}{\indsub{A}{\mu}},\rs{aB}{\indsub{A}{\Lambda}'}$, 
and of having nonlinear couplings that involve the skew part of 
the fields 
$\invsodder{[b}{BB'}\h{a]BB'}{\indsub{A}{\mu}}$. 

Finally, it is important to remark that, 
as in \Ref{Annalspaper,Henneaux2},
no conditions are assumed or required on the possibilities 
allowed for the form of the commutators of the deformed gauge symmetries
in the framework here. 
However, through the condition of gauge invariance of the Lagrangian,
closure of the deformed gauge symmetries on the solution space of
the deformed field equations will be seen to arise order by order, 
stemming from the fact that the abelian gauge symmetries 
generate all of the gauge freedom in the solutions of 
the linear field equations. 
Any deformation therefore automatically determines 
an associated infinitesimal gauge group structure.

\subsection{ Deformation results }

The solutions of the determining equations \eqrefs{liedereq}{liedercommeq}
can be obtained by the methods used for the deformation analysis
in \Ref{Annalspaper,AMSpaper}. 
We now outline the steps for the corresponding analysis here.
(See \Ref{thesis} for more details.)
To begin, 
the first order parts of all allowed deformations 
are found by solving 
the 0th-order part of 
the Lie derivative commutator equations \eqref{liedercommeq}
and 1st-order part of 
the Lie derivative equation \eqref{liedereq}
for, respectively, 
the linear terms in the gauge symmetries
$\varh{}{1}{aBB'}{\indsub{A}{\mu}},\varrs{}{1}{aB}{\indsub{A}{\Lambda}'}$
and quadratic terms in the field equations
$\Etop{2}{aBB'}{\indsub{A}{\mu}}{h},\Etop{2}{aB}{\indsub{A}{\Lambda}'}{\psi}$.
Calculation of the gauge symmetry commutators
$\var{0}{1}(\var{1}{2}\h{aBB'}{\indsub{A}{\mu}})
- \var{0}{2}(\var{1}{1}\h{aBB'}{\indsub{A}{\mu}})$
and 
$\var{0}{1}(\var{1}{2}\rs{aB}{\indsub{A}{\Lambda}'})
- \var{0}{2}(\var{1}{1}\rs{aB}{\indsub{A}{\Lambda}'})$
then determines the lowest-order part of 
the infinitesimal gauge group structure
$\varcomm{0}{1}{2}=\var{0}{3}$. 
Closure of this gauge group structure at the next lowest order
is derived from equations given by 
the 1st-order part of 
the Lie derivative commutator equations \eqref{liedercommeq}
minus the 1st-order part of 
the Lie derivative equation \eqref{liedereq} 
for the commutator gauge symmetries, 
where the field variables are taken to satisfy 
the linear field equations, 
$\Etop{1}{ab}{\indsub{A}{\mu}}{h} = \Etop{1}{aB}{\indsub{A}{\Lambda}'}{\psi}
=0$. 
When the gauge symmetry variables are taken to be rigid,
$\vect{a}{\indsub{A}{\mu}}=\const$, 
$\spin{B}{\indsub{A}{\Lambda}'}=\const$, 
$\lorspin{AB}{\indsub{A}{\mu}} =0$, 
so that 
$\var{0}{}\h{aBB'}{\indsub{A}{\mu}}= \var{0}{}\rs{aB}{\indsub{A}{\Lambda}'}
=0$,
the resulting Lie derivative equations 
are seen to impose integrability conditions 
on the first-order parts of the deformations. 
In particular, 
algebraic conditions arise on the coupling constants in 
the linear terms in the gauge symmetries
and quadratic terms in the field equations. 
These conditions are necessary (and, in fact, sufficient)
to allow solving for 
the quadratic terms in the gauge symmetries
$\varh{}{2}{aBB'}{\indsub{A}{\mu}},\varrs{}{2}{aB}{\indsub{A}{\Lambda}'}$
and cubic terms in the field equations
$\Etop{3}{aBB'}{\indsub{A}{\mu}}{h},\Etop{3}{aB}{\indsub{A}{\Lambda}'}{\psi}$
from, respectively, 
the 1st-order part of 
the Lie derivative commutator equations \eqref{liedercommeq}
and 2nd-order part of the Lie derivative equation \eqref{liedereq}
(with the gauge symmetry variables no longer being rigid
and the field variables no longer satisfying 
the linear field equations),
determining the second order parts of all allowed deformations.
Last, 
uniqueness of the second and higher order parts of the deformations 
is considered by an induction argument. 
Let $\Delta\varh{}{k}{aBB'}{\indsub{A}{\mu}}$,
$\Delta\varrs{}{k}{aB}{\indsub{A}{\Lambda}'}$, 
$\Delta\Etop{k+1}{aBB'}{\indsub{A}{\mu}}{h}$,
$\Delta\Etop{k+1}{aB}{\indsub{A}{\Lambda}'}{\psi}$
denote the difference of any two deformations that agree up to 
some finite order $1\le k\le \ell$.
These terms are shown to vanish at order $k=\ell+1$ by solving 
the $\ell$th-order part of 
the Lie derivative commutator equation \eqref{liedercommeq}
and the $\ell+1$st-order part of 
the Lie derivative equation \eqref{liedereq}. 
Hence it follows that any such deformations 
agree to all orders. 

To state the main results of this analysis,
it is convenient mathematically to 
view the set of ordinary \gron/ fields 
$\h{aBB'}{\indsub{A}{\mu}}$
as being an equivalent single \gron/ field $\algh{aBB'}{}$
possessing the internal structure of a real vector space 
$\Rnum{}^n\otimes\X$,
and similarly to view the set of ordinary \grino/ fields 
$\rs{aB}{\indsub{A}{\Lambda}'}$
as being an equivalent single \grino/ field $\algrs{aB}{}$
possessing the internal structure of a complexified vector space 
$\Rnum{}^{n'}\otimes\Y$. 

\Proclaim{ Theorem~1. }{
All non-higher-derivative deformations of
the linear abelian gauge theory \eqsref{linheq}{quadL}
for a set of ordinary \gron/ and \grino/ fields 
are (up to field redefinitions) equivalent to 
the nonlinear gauge theory of 
an algebra-valued tetrad field 
$\algframe{a}{BB'}= \openone \invsodder{a}{BB'} + \algh{a}{BB'}$
and an algebra-valued \RS/ field
$\algweyl{a}{B}= \algrs{a}{B}$
given by the chiral generalization of \NSG/ theory 
based on a modified \Gra/ algebra
intertwined with charge conjugation. 
(Here $\invsodder{a}{BB'}$ is a flat tetrad, 
and $\openone$ is the unit element in the algebra.)
}

The algebra on the vector spaces $\Rnum{}^n\otimes\X,\Rnum{}^{n'}\otimes\Y$, 
which we denote $\sgA{}$, 
underlying this nonlinear theory 
is defined by multiplication structure constants 
(real-valued) $\a{\indsub{A}{\mu}}{\indsub{B}{\alpha}\indsub{C}{\beta}}$, 
and (complex-valued) 
$\l{\indsub{A}{\mu}}{\indsub{B}{\Omega}'\indsub{C}{\Gamma}'}$, 
$\n{\indsub{A}{\Lambda}'}{\indsub{C}{\Gamma}'\indsub{B}{\alpha}}$, 
$\m{\indsub{A}{\Lambda}'}{\indsub{B}{\alpha}\indsub{C}{\Gamma}'}$.
These constants satisfy the following linear and quadratic relations
\EQs
&& 
\a{\indsub{A}{\mu}}{\indsub{B}{\alpha}\indsub{C}{\beta}} 
= \a{\indsub{A}{\mu}}{\indsub{C}{\beta}\indsub{B}{\alpha}} ,
\label{arel}\\
&& 
\l{\indsub{A}{\mu}}{\indsub{B}{\Omega}'\indsub{C}{\Gamma}'} 
= - \cl{\indsub{A}{\mu}}{\indsub{C}{\Gamma}'\indsub{B}{\Omega}'} ,
\label{lrel}\\
&& 
\m{\indsub{A}{\Lambda}'}{\indsub{B}{\alpha}\indsub{C}{\Gamma}'} 
= \cn{\indsub{A}{\Lambda}'}{\indsub{C}{\Gamma'}\indsub{B}{\alpha}} ,
\label{nmrel}
\endEQs
and 
\EQs
&& 
\a{\indsub{A}{\mu}}{\indsub{B}{\alpha}\indsub{D}{\nu}} 
\a{\indsub{D}{\nu}}{\indsub{C}{\beta}\indsub{E}{\sigma}} 
= \a{\indsub{A}{\mu}}{\indsub{C}{\beta}\indsub{D}{\nu}} 
\a{\indsub{D}{\nu}}{\indsub{B}{\alpha}\indsub{E}{\sigma}} ,
\label{aarel}\\
&& 
\n{\indsub{A}{\Lambda}'}{\indsub{B}{\Omega}'\indsub{C}{\alpha}} 
\n{\indsub{B}{\Omega}'}{\indsub{D}{\Gamma}'\indsub{E}{\mu}}
= \n{\indsub{A}{\Lambda}'}{\indsub{D}{\Gamma}'\indsub{B}{\nu}} 
\a{\indsub{B}{\nu}}{\indsub{C}{\alpha}\indsub{E}{\mu}} ,
\label{nnrel}\\
&& 
\a{\indsub{A}{\mu}}{\indsub{D}{\nu}\indsub{B}{\alpha}} 
\l{\indsub{D}{\nu}}{\indsub{C}{\Lambda}'\indsub{E}{\Gamma}'} 
= \l{\indsub{A}{\mu}}{\indsub{C}{\Lambda}'\indsub{B}{\Omega}'} 
\n{\indsub{B}{\Omega}'}{\indsub{E}{\Gamma}'\indsub{B}{\alpha}} ,
\label{alrel}\\
&& 
\n{\indsub{A}{\Lambda}'}{\indsub{C}{\Gamma}'\indsub{D}{\nu}} 
\l{\indsub{D}{\nu}}{\indsub{E}{\Sigma}'\indsub{B}{\Omega}'} 
= - \n{\indsub{A}{\Lambda}'}{\indsub{B}{\Omega}'\indsub{D}{\nu}} 
\l{\indsub{D}{\nu}}{\indsub{E}{\Sigma}'\indsub{C}{\Gamma}'} ,
\label{nlrel}
\endEQs
together with the additional relations
\EQs
&&
\q{\indsub{A}{\mu}\indsub{B}{\nu}}{}
= \q{\indsub{B}{\nu}\indsub{A}{\mu}}{} ,\quad
\q{\indsub{B}{\nu}\indsub{A}{\mu}}{}
\a{\indsub{A}{\mu}}{\indsub{C}{\alpha}\indsub{D}{\beta}}
= \q{\indsub{C}{\alpha}\indsub{A}{\mu}}{}
\a{\indsub{A}{\mu}}{\indsub{B}{\nu}\indsub{D}{\beta}} ,
\label{aqrel}\\
&&
\othq{\indsub{A}{\Lambda}'\indsub{B}{\Gamma}'}{}
= -\cothq{\indsub{B}{\Gamma}'\indsub{A}{\Lambda}'}{} ,\quad
\q{\indsub{B}{\nu}\indsub{A}{\mu}}{}
\l{\indsub{A}{\mu}}{\indsub{C}{\Lambda}'\indsub{D}{\Gamma}'}
= \othq{\indsub{C}{\Lambda}'\indsub{E}{\Omega}'}{}
\n{\indsub{E}{\Omega}'}{\indsub{D}{\Gamma}'\indsub{B}{\nu}} ,
\label{lnqrel}
\endEQs
where an overbar denotes complex conjugation. 
Note that, due to the equality \eqref{nmrel},
the structure constants 
$\m{\indsub{A}{\Lambda}'}{\indsub{B}{\alpha}\indsub{C}{\Gamma}'}$
will be suppressed hereafter. 

To conclude the main results, 
we note that Theorem~1 can be immediately specialized 
to the case of \gron/ fields alone. 

\Proclaim{ Theorem~2. }{
All non-higher-derivative deformations of
the linear abelian gauge theory 
\eqref{linheq}, \eqref{lindiffeosymm}, \eqref{linlorentzsymm}, \eqref{quadL}
for a set of ordinary \gron/ fields 
are (up to field redefinitions) equivalent to 
the nonlinear gauge theory given by the formulation of 
Einstein gravity theory for an algebra-valued metric field 
$\algmet{ab}{}= \openone \flat{ab} + \invsodder{(b}{BB'}\algh{a)BB'}{}$
using an even-\Gra/ algebra. 
}

In this nonlinear theory 
the even-\Gra/ algebra, which we denote $\gA$ 
on the vector space $\Rnum{}^n\otimes\X$, 
is defined by 
the multiplication structure constants 
(real-valued) $\a{\indsub{A}{\mu}}{\indsub{B}{\alpha}\indsub{C}{\beta}}$
satisfying the linear and quadratic relations
\eqref{arel}, \eqref{aarel}, \eqref{aqrel}.

\section{ Main results }
\label{results}

The results obtained in Theorem~1 for non-higher-derivative deformations
are now recast to give a classification of nonlinear gauge theories of
$n\ge 1$ \gron/ fields 
$\h{aBB'}{\mu}$ ($\mu=1,\ldots,n$)
and $n'\geq 1$ \grino/ fields 
$\rs{aB}{\Lambda}$ ($\Lambda=1,\ldots,n'$)
with formal internal multiplication rules. 
This will be accomplished by factorizing the algebra structure
on the vector spaces 
$\Rnum{}^n\otimes\X$ and $\Rnum{}^{n'}\otimes\Y$ 
defined by the coupling constants 
$\a{\indsub{A}{\mu}}{\indsub{B}{\alpha}\indsub{C}{\beta}}$, 
$\l{\indsub{A}{\mu}}{\indsub{B}{\Omega}'\indsub{C}{\Gamma}'}$,
$\n{\indsub{A}{\Lambda}'}{\indsub{C}{\Gamma}'\indsub{B}{\alpha}}$
associated with deformations
for the equivalent set of ordinary fields 
$\h{aBB'}{\indsub{A}{\mu}},\rs{aB}{\indsub{A'}{\Lambda}}$.

\subsection{ Algebraic structure }
\label{algebra}

To proceed, we consider the factorizations 
\EQ
\a{\indsub{A}{\mu}}{\indsub{B}{\alpha}\indsub{C}{\beta}} 
= \a{\mu}{\alpha\beta} \A{\ind{A}}{\ind{B}\ind{C}} ,\quad
\l{\indsub{A}{\mu}}{\indsub{B}{\Omega}'\indsub{C}{\Gamma}'} 
= \l{\mu}{\Omega\Gamma} \L{\ind{A}}{\ind{B}'\ind{C}'} ,\quad
\n{\indsub{A}{\Lambda}'}{\indsub{C}{\Gamma}'\indsub{B}{\alpha}}
= \n{\Lambda}{\Gamma\alpha} \N{\ind{A}'}{\ind{C}'\ind{B}} ,
\label{linfactor}
\endEQ
together with
\EQ
\q{\indsub{A}{\mu}\indsub{B}{\nu}}{}
= \q{\mu\nu}{} \Q{\ind{A}\ind{B}}{} ,\quad
\othq{\indsub{A}{\Lambda}'\indsub{B}{\Gamma}'}{}
= \othq{\Lambda\Gamma}{} \othQ{\ind{A}'\ind{B}'}{} ,
\label{linqfactor}
\endEQ
where $\a{\mu}{\alpha\beta}$, $\A{\ind{A}}{\ind{B}\ind{C}}$,
$\q{\mu\nu}{}$, $\Q{\ind{A}\ind{B}}{}$
are real constants, 
and $\l{\mu}{\Omega\Gamma}$, $\L{\ind{A}}{\ind{B}'\ind{C}'}$,
$\n{\Lambda}{\Gamma\alpha}$, $\N{\ind{A}'}{\ind{C}'\ind{B}}$,
$\othq{\Lambda\Gamma}{}$, $\othQ{\ind{A}'\ind{B}'}{}$
are complex constants. 
Mathematically, this corresponds to factorizing the algebra $\sgA{}$
defined by \eqsref{arel}{lnqrel}
on $\Rnum{}^n\otimes\X,\Rnum{}^{n'}\otimes\Y$, 
into an internal part on $\X\oplus\Y$, 
denoted $\multA{}$, 
with multiplication structure constants 
$\A{\ind{A}}{\ind{B}\ind{C}}$, 
$\L{\ind{A}}{\ind{B}'\ind{C}'}$,
$\N{\ind{A}'}{\ind{C}'\ind{B}}$,
and a $n+n'$ dimensional part on $\Rnum{}^n\oplus\Rnum{}^{n'}$, 
denoted $\hrsA{}$,
corresponding to the coupling constants
$\a{\mu}{\alpha\beta}$, 
$\l{\mu}{\Omega\Gamma}$,
$\n{\Lambda}{\Gamma\alpha}$,
for a set of $n$ \gron/ fields and $n'$ \grino/ fields. 
We now substitute the factorizations \eqrefs{linfactor}{linqfactor}
into the linear and quadratic algebraic relations \eqsref{arel}{lnqrel},
which correspondingly factorize into the following parts:
\EQs
&& 
\a{\mu}{\alpha\beta} 
= \lambda_1 \a{\mu}{\beta\alpha} ,\quad
\A{\ind{A}}{\ind{B}\ind{C}} 
= 1/\lambda_1 \A{\ind{A}}{\ind{C}\ind{B}} ,\quad
\lambda_1{}^2 =1 ,
\label{aArel}\\
&& 
\a{\mu}{\alpha\nu} \a{\nu}{\beta\sigma} 
= \lambda_2 \a{\mu}{\beta\nu} \a{\nu}{\alpha\sigma} ,\quad
\A{\ind{A}}{\ind{B}\ind{C}} \A{\ind{C}}{\ind{D}\ind{E}}
= 1/\lambda_2 \A{\ind{A}}{\ind{D}\ind{C}} \A{\ind{C}}{\ind{B}\ind{E}} ,\quad
\lambda_2{}^2 =1 ,
\label{aaAArel}
\endEQs
and
\EQs
&& 
\l{\mu}{\Lambda\Gamma} 
= \lambda_3 \cl{\mu}{\Gamma\Lambda} ,\quad
\L{\ind{A}}{\ind{B}'\ind{C}'} 
= -1/\lambda_3 \cL{\ind{A}}{\ind{C}'\ind{B}'} ,\quad
\lambda_3 {\bar\lambda}_3 =1, 
\label{lLrel}\\
&& 
\n{\Lambda}{\Gamma\nu} \l{\nu}{\Sigma\Omega} 
= \lambda_4 \n{\Lambda}{\Omega\nu} \l{\nu}{\Sigma\Gamma} ,\quad
\N{\ind{A}'}{\ind{B}'\ind{C}} \L{\ind{C}}{\ind{D}'\ind{E}'} 
= -1/\lambda_4 \N{\ind{A}'}{\ind{E}'\ind{C}} 
\L{\ind{C}}{\ind{D}'\ind{B}'} ,\quad
\lambda_4{}^2 =1 ,
\label{nlNLrel}\\
&& 
\a{\mu}{\nu\alpha} \l{\nu}{\Lambda\Gamma} 
= \lambda_5 \l{\mu}{\Lambda\Omega} \n{\Omega}{\Gamma\alpha} ,\quad
\A{\ind{A}}{\ind{B}\ind{C}} \L{\ind{B}}{\ind{D}'\ind{E}'}
= 1/\lambda_5 \L{\ind{A}}{\ind{D}'\ind{B}'} 
\N{\ind{B}'}{\ind{E}'\ind{C}} ,\quad
\lambda_5 \neq 0 ,
\label{alALrel}\\
&& 
\n{\Lambda}{\Omega\alpha} \n{\Omega}{\Gamma\beta}
= \lambda_6 \n{\Lambda}{\Gamma\nu} \a{\nu}{\alpha\beta} ,\quad
\N{\ind{A}'}{\ind{B}'\ind{C}} \N{\ind{B}'}{\ind{D}'\ind{E}} 
= 1/\lambda_6 \N{\ind{A}'}{\ind{D}'\ind{B}} 
\A{\ind{B}}{\ind{C}\ind{E}} ,\quad
\lambda_6 \neq 0 ,
\label{nnNNrel}
\endEQs
together with 
\EQs
&&
\q{\mu\nu}{} 
= \lambda_7 \q{\nu\mu}{} ,\quad
\Q{\ind{A}\ind{B}}{} 
= 1/\lambda_7 \Q{\ind{B}\ind{A}}{} ,\quad
\lambda_7{}^2 =1 ,
\label{qQrel}\\
&&
\q{\nu\mu}{} \a{\mu}{\alpha\beta} 
= \lambda_8 \q{\alpha\mu}{} \a{\mu}{\nu\beta} ,\quad
\Q{\ind{B}\ind{A}}{} \A{\ind{A}}{\ind{C}\ind{D}}
= 1/\lambda_8 \Q{\ind{C}\ind{A}}{} \A{\ind{A}}{\ind{B}\ind{D}} ,\quad
\lambda_8{}^2 =1 ,
\label{qaQArel}
\endEQs
and
\EQs
&&
\othq{\Lambda\Gamma}{} 
= \lambda_9 \cothq{\Gamma\Lambda}{} ,\quad
\othQ{\ind{A}'\ind{B}'}{} 
= -1/\lambda_9 \cothQ{\ind{B}'\ind{A}'}{} ,\quad
\lambda_9 {\bar\lambda}_9 =1 , 
\label{qQ'rel}\\
&&
\q{\nu\mu}{} \l{\mu}{\Lambda\Gamma} 
=  \lambda_{10} \q{\Lambda\Omega}{} \n{\Omega}{\Gamma\nu} ,\quad
\Q{\ind{B}\ind{A}}{} \L{\ind{A}}{\ind{C}'\ind{D}'} 
= 1/ \lambda_{10} \othQ{\ind{C}'\ind{A}'}{} 
\N{\ind{A}'}{\ind{D}'\ind{B}} ,\quad
\lambda_{10} \neq 0 ,
\label{qlQLrel}
\endEQs
for some constants $\lambda$. 
Note that the relation \eqref{aArel} implies $\a{\mu}{\alpha\beta}$ 
is either symmetric or antisymmetric in its lower indices. 
Compatibility of this index symmetry with the similar symmetry 
imposed by the relation \eqref{aaAArel} 
requires $\lambda_1=\lambda_2$
(or else $\a{\mu}{\alpha\nu} \a{\nu}{\beta\sigma}=0$
and $\A{\ind{A}}{\ind{B}\ind{C}} \A{\ind{C}}{\ind{D}\ind{E}}=0$). 
The same argument applied to relations \eqrefs{qQrel}{qaQArel}
leads to $\lambda_7=\lambda_8$. 
Likewise, in the case where $\l{\mu}{\Omega\Gamma}$ is taken to be real,
then compatibility of the resulting index symmetries 
imposed by \eqrefs{lLrel}{nlNLrel}
requires $\lambda_3=\lambda_4$
(or else $\n{\Lambda}{\Gamma\nu} \l{\nu}{\Sigma\Omega}=0$
and $\N{\ind{A}'}{\ind{B}'\ind{C}} \L{\ind{C}}{\ind{D}'\ind{E}'}=0$).

The main properties of the internal algebra on the vector spaces $\X,\Y$
and the field-coupling algebra on the vector spaces $\Rnum{}^n,\Rnum{}^{n'}$
defined by the algebraic equations \eqsref{aArel}{nnNNrel}
will now be summarized. 
In equations \eqrefs{aArel}{aaAArel},
the condition $\lambda_1=\lambda_2=\pm 1$
leads to two cases. 
For $\lambda_1=\lambda_2=1$, 
$\a{\mu}{\alpha\beta}$ defines multiplication structure constants of
a real, commutative, associative algebra 
$\Rnum{}^n\times\Rnum{}^n \into \Rnum{}^n$,
while $\A{\ind{A}}{\ind{B}\ind{C}}$ 
defines multiplication structure constants of
a real, anticommutative, anti-associative algebra 
$\X\times\X \into\X$. 
These algebras exchange properties for $\lambda_1=\lambda_2=-1$. 
The multiplication laws consist of
\EQs
&&
u_1 u_2 = \pm u_2 u_1 ,\quad
\varrho_1 \varrho_2 = \mp \varrho_2 \varrho_1 ,
\label{evencommrule}\\
&&
u_1 (u_2 u_3)= \pm (u_1 u_2) u_3 ,\quad
\varrho_1 (\varrho_2 \varrho_3)= \mp (\varrho_1 \varrho_2) \varrho_3 ,
\label{evenassocrule}
\endEQs
where $u$'s, $\varrho$'s stand for elements in $\Rnum^n,\X$. 

In equations \eqrefs{lLrel}{nlNLrel},
consider first the situation of real algebras, arising if 
$\l{\mu}{\Omega\Gamma} = \cl{\mu}{\Omega\Gamma}$,
$\L{\ind{A}}{\ind{B}'\ind{C}'} = \cL{\ind{A}}{\ind{B}'\ind{C}'}$
(which requires $\lambda_3=\lambda_4$). 
Then, the condition $\lambda_3=\lambda_4=\pm 1$
leads to two cases. 
For $\lambda_3=\lambda_4=1$, 
$\l{\mu}{\Omega\Gamma}$ and $\n{\Lambda}{\Gamma\alpha}$
define multiplication structure constants of
real, commutative, associative products 
$\Rnum{}^{n'}\times\Rnum{}^{n'} \into \Rnum{}^n$, 
$\Rnum{}^{n'}\times\Rnum{}^n \into \Rnum{}^{n'}$, 
while $\L{\ind{A}}{\ind{B}'\ind{C}'}$ and $\N{\ind{A}'}{\ind{C}'\ind{B}}$
define multiplication structure constants of
real, anticommutative, associative products
$\Y\times\Y \into \X$ and $\Y\times\X \into \Y$.
These products exchange properties for $\lambda_3=\lambda_4=-1$. 
The multiplication laws are given by 
\EQs
&&
o_1 o_2 = \pm o_2 o_1 ,\quad
\vartheta_1 \vartheta_2 = \mp \vartheta_2 \vartheta_1 , 
\label{realoddcommrule}\\
&&
o_1 (o_2 o_3)= \pm o_3 (o_2 o_1) = o_3 (o_1 o_2) ,\quad
\vartheta_1 (\vartheta_2 \vartheta_3)
= \mp \vartheta_3 (\vartheta_2 \vartheta_1)
= \vartheta_3 (\vartheta_1 \vartheta_2) , 
\label{oddassocrule}
\endEQs
where $o$'s, $\vartheta$'s stand for elements in $\Rnum^{n'},\Y$. 

The situation of complex algebras arises 
from equations \eqrefs{lLrel}{nlNLrel} if
$\l{\mu}{\Omega\Gamma} \neq \cl{\mu}{\Omega\Gamma}$,
$\L{\ind{A}}{\ind{B}'\ind{C}'} \neq \cL{\ind{A}}{\ind{B}'\ind{C}'}$.
There is no relationship required between 
$\lambda_3 =e^{i\theta}$ and $\lambda_4=\pm 1$
in this situation. 
These conditions then lead to 
two cases giving either associative or anti-associative multiplication rules
\eqref{evenassocrule}
and also to a one-parameter family of cases of 
generalized commutative-type multiplication rules
\EQ
o_1 o_2 = e^{\pm i\theta} \cc{\cc{o}_1 \cc{o}_2} ,\quad
\vartheta_1 \vartheta_2 
= -e^{\mp i\theta} \cc{\cc{\vartheta}_1 \cc{\vartheta}_2} , 
\label{oddcommrule}
\endEQ
in terms of a real constant $\theta$. 

Finally, from equations \eqrefs{alALrel}{nnNNrel},
the previous products satisfy 
generalized associative-type multiplication rules
\EQs
&&
(o_1 o_2) u_1 = \lambda o_1 (o_2 u_1) ,\quad
(\vartheta_1 \vartheta_2) \varrho_1 
= 1/\lambda \vartheta_1 (\vartheta_2 \varrho_1) , 
\label{oddevenassocrule}\\
&&
(o_1 u_1) u_2 = \lambda' o_1 (u_1 u_2) ,\quad
(\vartheta_1 \varrho_1) \varrho_2 
= 1/\lambda' \vartheta_1 (\varrho_1 \varrho_2) , 
\label{evenoddassocrule}
\endEQs
in terms of real constants $\lambda\neq 0,\lambda'\neq 0$.
Thus, 
strict associativity holds for the case $\lambda=\lambda'=1$,
whereas anti-associativity holds for the case $\lambda=\lambda'=-1$.
Other values of $\lambda,\lambda'$ yield generalized types of associativity. 

Hence the algebraic equations \eqsref{aArel}{nnNNrel}
together define an internal algebra $\multA{}$ on $\X\oplus \Y$
with formal multiplication rules \eqsref{evencommrule}{evenoddassocrule}
for the even and odd elements $\varrho$'s and $\vartheta$'s,
and a field-coupling algebra $\hrsA{}$ on $\Rnum{}^n\oplus\Rnum{}^{n'}$
with multiplication structure \eqsref{evencommrule}{evenoddassocrule}
for the even and odd elements $u$'s and $o$'s. 

The remaining algebraic equations \eqsref{qQrel}{qlQLrel}
determine an inner product on the internal algebra 
and on the field-coupling algebra,
as will be needed for construction of a scalar Lagrangian later. 
From equations \eqrefs{qQrel}{qaQArel}, 
the independent conditions $\lambda_7=\pm 1$, $\lambda_8=\pm 1$
yield four cases for the symmetry and invariance properties of
the inner products defined on $\Rnum{}^n,\X$. 
For $\lambda_7=\pm 1$, the inner products are respectively 
symmetric or antisymmetric
\EQ
<u_1,u_2> = \pm <u_2,u_1> ,\quad
<\varrho_1,\varrho_2> = \pm <\varrho_2,\varrho_1> , 
\label{evensymmmetric}
\endEQ
and for $\lambda_8=\pm 1$, invariant or anti-invariant
\EQ
<u_1,u_2 u_3> = \pm <u_2,u_1 u_3> ,\quad
<\varrho_1,\varrho_2 \varrho_3> = \pm <\varrho_2,\varrho_1 \varrho_3> . 
\label{eveninvmetric}
\endEQ
Similarly, from equations \eqrefs{qQ'rel}{qlQLrel},
the inner products defined on $\Rnum{}^{n'},\Y$
satisfy a two-parameter family of cases of symmetry properties
\EQ
<o_1,o_2> = e^{\pm i\theta} \cc{<\cc{o}_2,\cc{o}_1>} ,\quad
<\vartheta_1,\vartheta_2> 
= -e^{\mp i\theta} \cc{<\cc{\vartheta}_2,\cc{\vartheta}_1>} , 
\label{oddsymmmetric}
\endEQ
and invariance properties 
\EQ
<u_1,o_1 o_2> = \lambda <o_1, o_2 u_1> ,\quad
<\varrho_1, \vartheta_1 \vartheta_2> 
= 1/\lambda <\vartheta_1, \vartheta_2 \varrho_1> , 
\label{oddinvmetric}
\endEQ
in terms of a real constant $\theta$ and complex constant $\lambda\neq 0$. 
In the situation of real inner products, 
which arise if 
$\othq{\Lambda\Gamma}{} = \cothq{\Lambda\Gamma}{}$, 
$\othQ{\ind{A}'\ind{B}'}{} = \cothQ{\ind{A}'\ind{B}'}{}$, 
the symmetry property \eqref{oddsymmmetric} reduces to 
strict symmetry or antisymmetry
\EQ
<o_1,o_2> = \pm \cc{<\cc{o}_2,\cc{o}_1>} ,\quad
<\vartheta_1,\vartheta_2> 
= \mp \cc{<\cc{\vartheta}_2,\cc{\vartheta}_1>} 
\label{realoddsymmmetric}
\endEQ
while the invariance property \eqref{oddinvmetric}
is restricted to $\lambda=\bar\lambda$.

\subsection{ Nonlinear gauge theories of \gron/ fields }
\label{gravitonth}

To begin, consider for \gron/ fields alone 
the factorization of the even subalgebra of $\sgA{}$, 
\EQ
\hrsA{spin2} \otimes \multA{spin2} \equiv \gA
\label{evenfactorize}
\endEQ
defining a field-coupling algebra $\hrsA{spin2}$
for a set of $n\geq 1$ \gron/ fields 
$\h{aBB'}{\mu}$ 
whose formal multiplication rules are given by 
an internal algebra $\multA{spin2}$. 
This factorization combined with the deformation results in Theorem~2
yields a classification of nonlinear gauge theories of
\gron/ fields $\h{aBB'}{\mu}$ ($\mu=1,\ldots,n$)
with formal multiplication rules. 

\Proclaim{ Theorem~3. }{
All non-higher-derivative gauge theories of
a nonlinearly coupled set of $n\geq 1$ \gron/ fields $\h{aBB'}{\mu}$ 
using formal multiplication rules 
are (up to field redefinitions) given by 
the formulation of Einstein gravity theory 
for a single matrix-algebra valued metric field 
$\g{ab\indsub{B}{\nu}}{\indsub{A}{\mu}}
= \id{\indsub{B}{\nu}}{\indsub{A}{\mu}} \flat{ab}
+ \a{\indsub{A}{\mu}}{\indsub{B}{\nu}\indsub{C}{\sigma}} 
\invsodder{(b}{BB'} \h{a)BB'}{\indsub{C}{\sigma}}$
such that the algebra is a factorization of 
an associative, commutative algebra 
\eqref{evenfactorize}
as given by relations \eqrefs{aArel}{aaAArel}. 
The coupling constants and multiplication rules for $\h{aBB'}{\mu}$ 
are given by the algebras $\hrsA{spin2},\multA{spin2}$
in this factorization. 
}

These nonlinear \gron/ gauge theories for $\h{aBB'}{\mu}$ ($\mu=1,\ldots,n$)
are most naturally constructed in terms of metric tensor fields 
\EQ
\g{ab\nu}{\mu} 
= \a{\mu}{\nu\sigma} \h{ab}{\sigma}
+ \id{\nu}{\mu} \flat{ab}
\label{gfield}
\endEQ
and Christoffel connection tensors 
\EQ
\conn{c}{ab}{\mu} 
= \invg{cd\mu}{\nu}( \der{(a} \h{b)d}{\nu} -\frac{1}{2} \der{d}\h{ab}{\nu} )
\endEQ
where $\invg{ab\mu}{\nu}$ is the inverse of $\g{ab\nu}{\mu}$
satisfying 
\EQ
\invg{ac\mu}{\nu} \g{ab\sigma}{\nu} = \id{b}{c} \id{\sigma}{\mu} . 
\endEQ
Introduce the Riemann curvature associated with $\conn{c}{ab}{\mu}$ by 
\EQ
\R{abc}{d\mu} 
= \der{[a} \conn{d}{b]c}{\mu}
+ \a{\mu}{\alpha\beta} \conn{e}{c[a}{\alpha} \conn{d}{b]e}{\beta} ,
\endEQ
along with the Ricci tensor and scalar curvature 
\EQ
\R{ab}{\mu} = \R{acb}{c\mu} ,\quad
\R{}{\mu} = \invg{ab\mu}{\nu} \R{ab}{\nu} . 
\label{gR}
\endEQ
Then the field equations for $\h{ab}{\mu}$ are given by 
the vacuum Einstein tensor 
\EQ
\G{ab}{\mu} = \R{ab}{\mu} -\frac{1}{2} \g{ab\nu}{\mu} \R{}{\nu} =0 . 
\label{gheq}
\endEQ
The gauge invariance on solutions $\h{ab}{\mu}$ is given by 
a general-covariance symmetry 
\EQ
\delta_\xi\h{ab}{\mu} 
= \vect{}{c\sigma} \der{c}\g{ab\sigma}{\mu} 
+ 2\der{(a}\vect{}{c\sigma} \g{b)c\sigma}{\mu} 
\label{gdiffeosymm}
\endEQ
and also, trivially, by a local Lorentz symmetry 
\EQ
\delta_\chi\h{ab}{\mu} =0
\label{glorsymm}
\endEQ
since the skew part of $\h{aBB'}{\mu}$ does not enter $\h{ab}{\mu}$.
A gauge invariant Lagrangian is readily formulated 
by considering the densitized scalar curvature
\EQ
\lagr{\mu}{spin2} = \detg{\mu}{\nu} \R{}{\nu}
\label{gLagr}
\endEQ
where $\detg{\mu}{\nu}$ is the metric volume density given by 
\EQ
\detg{\mu}{\nu} \detg{\nu}{\sigma}  
= \g{am\alpha}{\mu} \g{bn\beta}{\alpha} 
\g{ck\gamma}{\beta} \g{dl\sigma}{\gamma}
\invvol{abcd}\invvol{mnkl} .
\endEQ
Under the general-covariance symmetry \eqref{gdiffeosymm}, 
$\lagr{\mu}{spin2}$ is invariant 
(to within a total divergence $\der{a} \S{a\mu}$)
and in addition is trivially invariant with respect to 
the local Lorentz symmetry \eqref{glorsymm}.
Moreover, by variation of $\h{ab}{\mu}$,
$\lagr{\mu}{spin2}$ yields the field equations \eqref{gheq},
such that 
$\delta_h \lagr{\mu}{spin2} 
= \a{\mu}{\nu\sigma} \G{}{ab\nu} \delta\h{ab}{\sigma} +$ a total divergence. 
Construction of an equivalent scalar Lagrangian will be addressed shortly.
It is important to emphasize here that,
in the expressions for the Lagrangian, field equations, and gauge symmetries,
all products of the fields $\h{ab}{\mu}$ 
involve formal multiplication given by the rules of the internal algebra 
$\multA{spin2}$
(so any change in the order or arrangement of the fields 
requires using these formal rules). 
As a consequence, mathematically, 
these expressions are not real-valued but rather are regarded formally as
taking values in $\multA{spin2}$. 

From the algebraic results in \secref{algebra},
we see that there are only two types allowed for 
the multiplication rules $\multA{spin2}$
and the accompanying coupling constants $\hrsA{spin2}$
in the nonlinear theories \eqsref{gfield}{gLagr}. 
One type is, obviously, 
where $\hrsA{spin2}$ is an arbitrary $n$ dimensional 
associative, commutative algebra
and $\multA{spin2}$ consists of standard multiplication
so each $\h{ab}{\mu}$ is formally a commuting field. 
In this situation, 
a simple case for $\multA{spin2}$ is just given by 
the algebra of real numbers, $\Rnum$,
and correspondingly, 
$\h{ab}{\mu}$ is a set of ordinary (\ie/ real-valued) fields. 
The nonlinear theories \eqsref{gfield}{gLagr} 
thereby are precisely the same as the multi-graviton theories
first found by Cutler and Wald. 
This is most easily seen by 
employing a unit element $\unit{\mu}$ in $\hrsA{spin2}$
(appending one if none exists \cite{Wald}). 
Then, through the relations 
$\a{\mu}{\nu\sigma} \unit{\sigma} = \id{\nu}{\mu}$
and $\g{ab\nu}{\mu} = \a{\mu}{\nu\sigma} \g{ab}{\sigma}$,
the nonlinear theories \eqsref{gfield}{gLagr} 
simplify to Einstein gravity theory 
for the $\hrsA{spin2}$-valued metric tensor 
\EQ
\algmet{ab}{}= \openone \flat{ab} + \invsodder{(b}{BB'}\algh{a)BB'}{}
\label{algmetric}
\endEQ
with $\multA{spin2}=\Rnum$. 

The other allowed type is instead that 
$\hrsA{spin2}$ is an arbitrary $n$ dimensional 
anticommutative, anti-associative algebra
and $\multA{spin2}$ comprises formal rules of 
anticommutative, anti-associative multiplication. 
In this situation 
the fields $\h{ab}{\mu}$ are each formally anticommuting 
and obey an anti-associativity relation for formal products 
with three (or more) fields,
\EQ
\varrho_1 \varrho_2 = - \varrho_2 \varrho_1 ,\quad
\varrho_1 (\varrho_2 \varrho_3)= - (\varrho_1 \varrho_2) \varrho_3 ,
\label{gnewmultrule}
\endEQ
where $\varrho$'s stand for $\h{ab}{\mu}$ 
or products of any number of $\h{ab}{\mu}$'s. 
An anticommutative, anti-associative field-coupling algebra
$\hrsA{spin2}$ is characterized by structure constants 
with the antisymmetry properties
\EQ
\a{\mu}{(\alpha\beta)} =0 ,\quad
\a{\mu}{\nu(\alpha} \a{\nu}{\beta)\sigma} =0 .
\label{gnewcouplingrule}
\endEQ
A simple nontrivial example is given by 
$\a{4}{1\ 2} \neq 0$, $\a{5}{1\ 3} \neq 0$, $\a{6}{2\ 3} \neq 0$, 
$\a{7}{3\ 4} = \a{7}{1\ 6} = -\a{7}{2\ 5} \neq 0$
(all others zero, taking into account antisymmetry) 
with $n=7$,
producing a seven-dimensional anticommutative, anti-associative algebra 
$\hrsA{spin2}$. 
(This example arises by forming all possible 
anticommutative, anti-associative products of three generators:
$u_1,u_2,u_3,
u_{12}\equiv u_1 u_2 =-u_2 u_1,
u_{13}\equiv u_1 u_3 = -u_3 u_1,
u_{23}\equiv u_2 u_3 = -u_3 u_2,
u_{123}\equiv 
u_1 u_{23} = -u_{23} u_1= -u_2 u_{13}= u_{13} u_2= u_3 u_{12} =-u_{12} u_3$,
and all other products equal to zero.)
These antisymmetric structure constants $\a{\mu}{\alpha\beta}$ 
yield a nonlinear multi-graviton theory \eqsref{gfield}{gLagr} 
for a set of seven coupled anticommuting \gron/ fields 
$\h{aBB'}{\mu}$ ($\mu=1,\ldots,7$). 
As illustrated by this example,
the anticommuting nature (and anti-associativity) of $\h{aBB'}{\mu}$ here 
is tied to the number $n$ of \gron/ fields being at least two. 

Moreover, in contrast to the commutative, associative type of algebra
for $\hrsA{spin2}$ in the Cutler-Wald multi-graviton theories, 
the new multi-graviton theories presented here 
for an anticommutative, anti-associative type of algebra $\hrsA{spin2}$
do not have a formulation in terms of 
an $\hrsA{spin2}$-valued metric tensor \eqref{algmetric},
since no anticommutative algebra can admit a unit element. 

Finally, 
we turn to constructing a scalar Lagrangian 
for these multi-graviton theories. 
The construction makes use of an invariant inner product 
on the field-coupling algebra $\hrsA{spin2}$ as follows. 

Consider, first of all, 
the case of an associative, commutative algebra $\hrsA{spin2}$ 
on $\Rnum{}^n$, 
with structure constants $\a{\mu}{\alpha\beta}$. 
For simplicity, 
it is sufficient to assume $\a{\mu}{\alpha\beta}$ is irreducible
(namely, that the algebra is not a direct sum of two or more 
nontrivial subalgebras).
An invariant symmetric inner product with nondegenerate components 
$\q{\mu\nu}{}= \q{(\mu\nu)}{}$
is characterized by the invariance relation 
\EQ
\q{\mu[\nu}{} \a{\mu}{\alpha]\beta} =0 .
\label{gmetricrule}
\endEQ
Now, suppose $\a{\mu}{\alpha\beta}$ possesses a unit element $\unit{\mu}$.
Then the relation \eqref{gmetricrule} shows that 
\EQ
\q{\nu\sigma}{} = \dualq{\mu}{} \a{\mu}{\nu\sigma}
\label{gunitmetric}
\endEQ
where $\dualq{\mu}{} = \q{\mu\nu}{} \unit{\nu}$ 
represents the dual of $\unit{\nu}$ with respect to the inner product. 
Consequently, 
any inner product is obtained by fixing some constants $\dualq{\mu}{}$
such that the components \eqref{gunitmetric} are nondegenerate. 
If there is a nilpotent element in the algebra,
then nondegeneracy holds if and only if 
the most nilpotent ideal is one-dimensional 
and $\dualq{\mu}{}$ is chosen to be nonvanishing on a most nilpotent element.
An example of such algebras is any even-\Gra/ algebra, $\evenG{unit}$.
Alternatively, if there are no nilpotent elements in the algebra,
then due to the irreducibility assumption \cite{Wald}, 
the algebra is necessarily isomorphic to $\Rnum$ or $\Cnum$
and then it is sufficient to choose $\dualq{\mu}{}$ 
to be nonvanishing on the unit element. 

\Proclaim{ Proposition~3. }{
For an associative, commutative algebra $\hrsA{spin2}$
with a unit element 
and with a most nilpotent ideal of at most one dimension, 
a scalar Lagrangian for the nonlinear theories \eqsref{gfield}{gLagr} 
using formal associative, commutative multiplication rules
for $\h{ab}{\mu}$ 
is obtained simply by taking
\EQ
\lagr{}{spin2} = \dualq{\mu}{} \lagr{\mu}{spin2}
\label{gscalarL}
\endEQ
where $\dualq{\mu}{}$ projects onto 
the most nilpotent element or, if none, the unit element. 
}

This Lagrangian is invariant (to within a total divergence)
under the gauge symmetries \eqrefs{gdiffeosymm}{glorsymm}
and yields the field equations \eqref{gheq}
by variation of $\h{ab}{\mu}$. 

Next suppose $\hrsA{spin2}$ is an associative, commutative algebra
that is completely nilpotent. 
In this case, for any constants $\dualq{\mu}{}$, 
clearly the components \eqref{gunitmetric} are degenerate
on the most nilpotent ideal
and so the scalar Lagrangian \eqref{gscalarL} breaks down
as it does not contain (at least) the field $\dualq{\mu}{}\h{ab}{\mu}$.
Of course, 
the relation \eqref{gunitmetric} is merely sufficient,
but not necessary, 
for $\q{\nu\sigma}{}$ to satisfy the invariance property \eqref{gmetricrule}
when $\hrsA{spin2}$ does not possess a unit element. 
However, 
for the example of nilpotent even-\Gra/ algebras, $\evenG{nilpotent}$,
any invariant $\q{\nu\sigma}{}$ is degenerate 
on the most nilpotent ideal.
(For instance, 
consider the even-\Gra/ algebra with three nilpotent generators:
$u_1,u_2,u_3,
u_{12}\equiv u_1 u_2, u_{13}\equiv u_1 u_3, u_{23}\equiv u_2 u_3,
u_{123}\equiv u_1 u_2 u_3$. 
Since the square of any element is zero,
the invariance property of $\q{\nu\sigma}{}$ applied to 
$<u_{123},u> = <u_1,u u_{23}> = <u_2,u u_{13}> = <u_3,u u_{12}>$ 
forces this to vanish for all elements $u$.)
An example on the other hand 
where an invariant nondegenerate $\q{\nu\sigma}{}$ exists 
is any nilpotent monogenic associative, commutative algebra, $\V{p}$.
(Namely, consider an associative, commutative algebra 
that is generated by powers of a single element $u$ such that $u^{p+1}=0$.
Then the inner product defined by $<u^j,u^k>=\id{p}{j+k}$ 
is, clearly, invariant and nondegenerate on the entire algebra.)
In this case, 
a scalar Lagrangian is constructed by the following procedure.

\Proclaim{ Proposition~4. }{
For a completely nilpotent associative, commutative algebra $\hrsA{spin2}$
with an invariant nondegenerate inner product, 
a scalar Lagrangian 
for the nonlinear theories \eqsref{gfield}{gLagr} 
using formal associative, commutative multiplication rules
for $\h{ab}{\mu}$ 
is given in terms of components $\q{\mu\nu}{}$ of the inner product 
by 
\EQ
\lagr{}{spin2} = \q{\mu\nu}{} \lagr{\mu\nu}{}
\label{gnilpscalarL}
\endEQ
where $\lagr{\mu\nu}{}$ is the coefficient of
$\a{\sigma}{\mu\nu}$ in $\lagr{\sigma}{spin2}$. 
}

This construction works because, 
if we split 
$\lagr{\sigma}{spin2} = \a{\sigma}{\mu\nu} \lagr{\mu\nu}{} +$
linear $\h{ab}{\sigma}$ terms, 
the linear terms are found to be a total divergence
and so $\q{\mu\nu}{} \lagr{\mu\nu}{}$ retains
the gauge symmetries and field equations of $\lagr{\sigma}{spin2}$
due to the invariance property of $\q{\mu\nu}{}$. 

Last, 
consider the case of an anticommutative, anti-associative algebra 
$\hrsA{spin2}$ on $\Rnum{}^n$, 
with structure constants $\a{\mu}{\alpha\beta}$. 
An invariant antisymmetric inner product with nondegenerate components 
$\q{\mu\nu}{}= \q{[\mu\nu]}{}$
is characterized by the invariance relation 
\EQ
\q{\mu(\nu}{} \a{\mu}{\alpha)\beta} =0 .
\label{gskewmetricrule}
\endEQ
Observe, here, that due to anticommutativity,
the square of any element in $\hrsA{spin2}$ vanishes
and hence the algebra is completely nilpotent. 
Consequently, 
although the relation \eqref{gunitmetric}
continues to yield an invariant inner product, 
it is degenerate on all most nilpotent elements. 
Moreover, 
an argument similar to the one for the example of
a nilpotent even-\Gra/ algebra $\evenG{nilpotent}$ considered previously 
shows that any invariant inner product for
an anticommutative, anti-associative algebra 
generated analogously to $\evenG{nilpotent}$
must always be degenerate. 
However, if an extra element is appended in a suitable manner,
then an invariant nondegenerate inner product exists 
on the enlarged algebra. 
(In particular, 
for the example of 
the seven-dimensional anticommutative, anti-associative algebra 
with three generators $u_1,u_2,u_3$ given earlier, 
the nilpotency property $u_1{}^2=u_2{}^2=u_3{}^2=0$
combined with the invariance property 
$<u_{123},u> = <u_1,u u_{23}> = <u_2,u_{13} u> = <u_3,u u_{12}>$
forces this inner product to vanish for all elements $u$.
But, by appending an extra element $u_0$
and imposing $u_0 u_{123}=0$, 
it follows that
$<u_{123},u_0>= <u_1,u_0 u_{23}>= <u_2,u_{13} u_0>= <u_3,u_0 u_{12}>\neq 0$
determines an invariant nondegenerate inner product.)
More generally, 
if we mod out with respect to the most nilpotent ideal 
in an anticommutative, anti-associative algebra 
formed from any number of generators, 
then the quotient algebra admits an invariant nondegenerate inner product.

The construction of a scalar Lagrangian in this case 
now parallels Proposition~4.

\Proclaim{ Proposition~5. }{
For an anticommutative, anti-associative algebra $\hrsA{spin2}$
with an invariant nondegenerate inner product, 
a scalar Lagrangian 
for the nonlinear theories \eqsref{gfield}{gLagr} 
using formal anticommutative, anti-associative multiplication rules 
for $\h{ab}{\mu}$ 
is given by \eqref{gnilpscalarL}
in terms of components $\q{\mu\nu}{}$ of the inner product. 
}

In all cases, 
the Lagrangians \eqref{gLagr}, \eqref{gscalarL}, \eqref{gnilpscalarL}
are polynomial in $\h{ab}{\mu}$ (and its derivatives)
when, and only when, the field-coupling algebra $\hrsA{spin2}$
is completely nilpotent. 
Moreover, in this situation these Lagrangians depend essentially on 
the background flat tetrad $\invsodder{}{aBB'}$. 
Indeed, the nonlinear theories \eqsref{gfield}{gLagr} 
are not independent of this flat tetrad 
unless the field-coupling algebra possesses a unit element. 

As a final remark 
it is worth noting that the inner products on $\hrsA{spin2}$
underlying these scalar Lagrangians are of indefinite sign
\cite{Anco-Wald,thesis,Henneaux2}
whenever the field-coupling algebra is nontrivial
(in particular, if the most nilpotent ideal is a proper subalgebra). 
Consequently, 
the canonical stress-energy tensor derived from the scalar Lagrangian 
associated to the set of coupled fields $\h{ab}{\mu}$
lacks any formal positivity properties,
in contrast to the well-known dominant energy property
of the stress-energy tensor in the case of a single field $\h{ab}{}$.

\subsection{ Supersymmetric extensions }

It is straightforward to construct supersymmetric extensions of 
the nonlinear multi-graviton theories \eqsref{gfield}{gLagr},
starting from a factorization of the algebra 
$\sgA{} =\hrsA{} \otimes \multA{}$
to obtain a field-coupling algebra $\hrsA{}$ for a set of 
$n\geq 1$ \gron/ fields 
$\h{aBB'}{\mu}$ ($\mu=1,\ldots,n$)
and $n'\geq 1$ \grino/ fields 
$\rs{aB}{\Lambda}$ ($\Lambda=1,\ldots,n'$)
whose formal multiplication rules are given by 
an internal algebra $\multA{}$.

From the algebraic results in \secref{algebra},
we see that the allowed multiplication rules $\multA{}$
and accompanying coupling constants $\hrsA{}$
comprise a rich variety of types. 
First of all, note the algebras possess a semidirect product structure
stemming from a natural grading into even and odd parts,
$\hrsA{even/odd}$ and $\multA{even/odd}$,
which correspond respectively to 
the \gron/ fields (assigned even-grading)
and the \grino/ fields (assigned odd-grading).
The even parts $\hrsA{even}$ and $\multA{even}$
are given by the two types of algebras $\gA$
discussed in \secref{gravitonth}
for the case of \gron/ fields alone. 
The odd parts $\hrsA{odd}$ and $\multA{odd}$
involve the product structure 
$\sgA{odd} \times \sgA{odd}$ into $\sgA{even}$,
called odd multiplication, 
and $\sgA{odd}\times\sgA{even}$ into $\sgA{odd}$,
called even-odd multiplication, 
which defines the semidirect product of $\sgA{even}$ with $\sgA{odd}$.
In addition, $\sgA{odd}$ is allowed to be complexified
whereas $\sgA{even}$ is necessarily real,
with complex conjugation representing charge conjugation 
on the \gron/ and \grino/ fields. 

In the case of real algebras $\sgA{}$,
the only allowed types of odd multiplication are
associative and either anticommutative or commutative,
while the allowed even-odd multiplication exhibits 
a two-parameter type of generalized associativity. 
The case of complexified algebras $\sgA{}$ is richer in allowed types,
due to intertwining of complex conjugation with 
both odd multiplication and even-odd multiplication. 
This will not be explored in further detail here. 

To proceed, 
the supersymmetric extension of the nonlinear theories \eqsref{gfield}{gLagr}
for a set of fields 
$\h{aBB'}{\mu}$ ($\mu=1,\ldots,n$)
and $\rs{aB}{\Lambda}$ ($\Lambda=1,\ldots,n'$)
with coupling constants given by $\hrsA{}$
and using formal internal multiplication rules of $\multA{}$
is based on the following main result. 

\Proclaim{ Theorem~4. }{
All non-higher-derivative gauge theories of
a nonlinearly coupled set of $n\geq 1$ \gron/ fields $\h{aBB'}{\mu}$ 
and $n'\geq 1$ \grino/ fields $\rs{aB}{\Lambda}$ 
using formal multiplication rules 
are (up to field redefinitions) given by 
the chiral generalization of \NSG/ theory \cite{PhysRevpaper}
for a single real, matrix-algebra valued tetrad field 
$\frame{}{aBB'\indsub{B}{\nu}}{\indsub{A}{\mu}}
= \id{\indsub{B}{\nu}}{\indsub{A}{\mu}} \invsodder{aBB'}{}
+ \a{\indsub{A}{\mu}}{\indsub{B}{\nu}\indsub{C}{\sigma}} 
\h{aBB'}{\indsub{C}{\sigma}}$
and a single conjugate-pair of 
complex, matrix-algebra valued vector-spinor fields
$\weyl{}{aB\indsub{B}{\nu}}{\indsub{A}{\Lambda}'}
= \n{\indsub{A}{\Lambda}'}{\indsub{C}{\Gamma}'\indsub{B}{\nu}}
\rs{aB}{\indsub{C}{\Gamma}'}$, 
$\cweyl{}{aB'\indsub{C}{\Gamma}'}{\indsub{A}{\mu}}
= \l{\indsub{A}{\mu}}{\indsub{B}{\Lambda}'\indsub{C}{\Gamma}'}
\crs{aB'}{\indsub{B}{\Lambda}'}$
such that the algebra is a factorization of 
a modified associative, graded-commutative algebra 
$\sgA{} = \hrsA{} \otimes \multA{}$
that intertwines with charge conjugation
as given by relations \eqsref{aArel}{nnNNrel}. 
The coupling constants and multiplication rules 
for $\h{aBB'}{\mu}$ and $\rs{aB}{\Lambda}$ 
are given by the algebras $\hrsA{},\multA{}$
in this factorization. 
}

The resulting nonlinear gauge theories 
for the fields $\h{aBB'}{\mu},\rs{aB}{\Lambda}$ 
($\mu=1,\ldots,n$, $\Lambda=1,\ldots,n'$)
are most naturally formulated in terms of 
vector-spinors
\EQ
\weyl{}{aB\nu}{\Lambda} 
= \n{\Lambda}{\Gamma\nu} \rs{aB}{\Gamma} ,\quad
\cweyl{}{aB'\Gamma}{\mu}
= \l{\mu}{\Lambda\Gamma} \crs{aB'}{\Lambda} ,
\label{sgweylfield}
\endEQ
and tetrads 
\EQ
\frame{}{aBB'\nu}{\mu} 
= \a{\mu}{\nu\sigma} \h{aBB'}{\sigma}
+ \id{\nu}{\mu} \invsodder{aBB'}{} ,\quad
\frame{}{aBB'\Gamma}{\Lambda} 
= \n{\Lambda}{\Gamma\sigma} \h{aBB'}{\sigma}
+ \id{\Gamma}{\Lambda} \invsodder{aBB'}{} ,
\label{sgefield}
\endEQ
along with Lorentz spin connections and curvatures
\EQ
\w{a}{CD\mu} =\w{a}{(CD)\mu} ,\quad
\R{ab}{CD\mu} 
= \der{[a} \w{b]}{CD\mu} 
+ \a{\mu}{\alpha\beta} \w{[a}{CE\alpha} \w{b]E}{D\beta}
\endEQ
determined by the torsion equation
\EQ
\der{[b} \frame{DD'}{c]\nu}{\mu} 
- \a{\mu}{\nu\sigma} \w{[b|B}{D\alpha} \frame{BD'}{|c]\alpha}{\sigma} 
+ c.c.
= - \cweyl{D'}{[b|\Gamma}{\mu} \weyl{D}{|c]\nu}{\Gamma} .
\label{torsion}
\endEQ
This equation can be solved for $\w{a}{CD\mu}$ 
by using the inverse tetrads which satisfy
\EQ
\frame{}{aCC'\sigma}{\nu} \invframe{}{aBB'\mu}{\nu} 
= \cross{B}{C}\cross{B'}{C'} \id{\sigma}{\mu} ,\quad
\frame{}{aCC'\Omega}{\Gamma} \invframe{}{aBB'\Lambda}{\Gamma} 
= \cross{B}{C}\cross{B'}{C'} \id{\Omega}{\Lambda} , 
\endEQ
where $\vol{BC}$ is the spin metric. 
The field equations for $\h{aBB'}{\mu}$ and $\rs{aB}{\Lambda}$ 
are given by 
the Einstein equation with \grino/ matter source
\EQ
\G{}{aBB'\mu} -\T{}{aBB'\mu} =0 ,
\label{sgheq}
\endEQ
and by the \RS/ equation
\EQ
\F{}{[bc|D}{\Gamma} \frame{DD'}{|a]\Gamma}{\Lambda} =0 ,
\label{sgrseq}
\endEQ
where
\EQ
\F{}{cdB}{\Lambda} 
= \der{[c} \rs{d]B}{\Lambda} 
+ \weyl{}{[d|E\sigma}{\Lambda} \w{|c]B}{E\sigma}
\endEQ
is the covariant \RS/ field strength,
and where 
\EQ
\G{}{aBB'\mu} 
= \frame{BB'}{b\nu}{\mu} 
( \R{}{ab\nu} -\frac{1}{2} \invg{ab\nu}{\sigma}\R{}{\sigma} )
\endEQ
is the spinorial Einstein tensor
for the metric fields 
\EQ
\g{ab\nu}{\mu} = \frame{}{aBB'\sigma}{\mu} \frame{BB'}{b\nu}{\sigma} ,
\label{sggfield}
\endEQ
with Ricci tensor $\R{}{ab\nu}$ and scalar curvature $\R{}{\sigma}$
given by \eqref{gR}, 
and 
\EQ
\T{}{aBB'\mu} 
= \frac{1}{2} i\vole{}{abcd\mu}{\nu}
\cweyl{B'}{b\Gamma}{\nu} \F{B}{cd}{\Gamma} + c.c. 
\endEQ
is the spinorial stress-energy tensor. 
Here 
\EQ
\vole{abcd\nu}{\mu}{} 
= \dete{\mu}{\nu} \vol{abcd} 
= 2i \frame{A}{[a|A'\alpha|}{\mu} \frame{A'}{b|B\beta|}{\alpha} 
\frame{B}{c|C'\gamma|}{\beta} \frame{C'}{d]A\nu}{\gamma}
\endEQ
is the tetrad volume tensor,
with 
\EQ
\dete{\mu}{\nu} = \detg{\mu}{\nu} .
\endEQ
Gauge invariance on solutions $\h{aBB'}{\mu}$ and $\rs{aB}{\Lambda}$
is given by 
a general-covariance symmetry 
\EQ
\delta_\xi\h{aBB'}{\mu} 
= \vect{}{c\sigma} \der{c}\frame{}{aBB'\sigma}{\mu} 
+ \der{a}\vect{}{c\sigma} \frame{}{cBB'\sigma}{\mu} ,\quad
\delta_\xi\rs{aB}{\Lambda}
= \vect{}{c\sigma} \der{c}\weyl{}{aB\sigma}{\Lambda}
+ \der{a}\vect{}{c\sigma} \weyl{}{cB\sigma}{\Lambda} ,
\label{sgdiffeosymm}
\endEQ
as well as a local Lorentz symmetry 
\EQ
\delta_\chi\h{aBB'}{\mu} 
= \lorspin{B}{C\nu} \frame{}{aCB'\nu}{\mu} + c.c. ,\quad
\delta_\chi\rs{aB}{\Lambda}
= \weyl{}{aC\sigma}{\Lambda} \lorspin{B}{C\sigma} ,
\label{sglorsymm}
\endEQ
and a \susy/ 
\EQ
\delta_\zeta\h{aBB'}{\mu}
= \cweyl{}{aB'\Gamma}{\mu} \spin{B}{\Gamma} + c.c. ,\quad
\delta_\zeta\rs{aB}{\Lambda}
= \der{a}\spin{B}{\Lambda} 
+ \n{\Lambda}{\Gamma\sigma} \spin{c}{\Gamma} \w{aB}{C\sigma} ,
\label{sgsusy}
\endEQ
for arbitrary vector fields $\vect{}{c\sigma}$
and arbitrary spinor fields 
$\spin{B}{\Gamma}$ and $\lorspin{BC}{\nu}= \lorspin{(BC)}{\nu}$.
A gauge invariant Lagrangian is obtained 
by considering the densitized sum of
a gravitational scalar curvature term
and matter current term 
\EQ
\lagr{\mu}{spin2,3/2} 
= \invvol{abcd} \frac{1}{4} i \Big( 
2(\frame{}{cCC'\nu}{\mu} \frame{C'}{Dd\sigma}{\nu}) 
\R{ab}{CD\sigma} 
+ \cweyl{}{aB\Gamma}{\mu} (\F{}{cdB}{\Omega} \frame{BB'}{b\Omega}{\Gamma}) 
\Big) + c.c.\ .
\label{sgLagr}
\endEQ
Under the symmetries \eqsref{sgdiffeosymm}{sgsusy}, 
$\lagr{\mu}{spin2,3/2}$ is invariant 
(to within a total divergence $\der{a} \S{a\mu}$). 
In addition, by variation of $\h{aBB'}{\mu}$ and $\rs{aB}{\Lambda}$,
$\lagr{\mu}{spin2,3/2}$ yields the field equations \eqrefs{sgheq}{sgrseq}. 
Note that 
all products of the fields $\h{aBB'}{\mu}$ and $\rs{aB}{\Lambda}$
in the Lagrangian, field equations, and gauge symmetries
involve formal multiplication rules given by $\multA{}$
(which must be used for reordering or rearranging of products). 
Consequently, 
these expressions are to be regarded formally as taking values in $\multA{}$.

When $\hrsA{}$ is an arbitrary $n+n'$ dimensional 
commutative, associative graded algebra
and $\multA{}$ consists of standard \Gra/ multiplication, 
the nonlinear theories \eqsref{sgweylfield}{sgLagr}
describe a $N=1$ supersymmetric extension of 
the multi-graviton gauge theories of Cutler and Wald,
where the $n>1$ \gron/ fields $\h{aBB'}{\mu}$
are each formally commuting
while the $n'>1$ \grino/ fields $\rs{aB}{\Lambda}$
are each formally anticommuting
(with all products obeying ordinary associativity). 
The odd-\Gra/ multiplication rules in this extension
can be modified to intertwine with charge conjugation,
so $\rs{aB}{\Lambda}$ is no longer strictly anticommuting
while $\h{aBB'}{\mu}$ remains commuting
(and products are no longer strictly associative) 
analogously to the single graviton and graviton case ($n=n'=1$)
discussed in \Ref{PhysRevpaper}. 
The field coupling constants can be generalized correspondingly,
such that commutativity and associativity 
with respect to $\hrsA{odd}$ hold only up to complex conjugation. 

A different extension of the Cutler-Wald multi-graviton theories
is given by using standard multiplication rules
for the \gron/ and \grino/ fields, 
corresponding to when $\multA{}$ is 
simply the algebra of real numbers, $\Rnum$,
while $\hrsA{}$ is now a real, graded-commutative, associative algebra,
whose structure constants 
$\a{\mu}{\alpha\beta}$, 
$\l{\mu}{\Omega\Gamma}$,
$\n{\Lambda}{\Gamma\alpha}$
are characterized by the relations 
\EQs
&&
\a{\mu}{[\alpha\beta]} =0 , 
\label{gcouplingrule}\\
&&
\l{\mu}{(\Omega\Gamma)} =0 ,\quad
\n{\Lambda}{(\Gamma|\nu|} \l{\nu}{\Sigma)\Omega} =0 , 
\label{sgcouplingrule}
\endEQs
together with
\EQ
\a{\mu}{\nu\alpha} \l{\nu}{\Lambda\Gamma} 
= \l{\mu}{\Lambda\Omega} \n{\Omega}{\Gamma\alpha} ,\quad
\n{\Lambda}{\Omega\alpha} \n{\Omega}{\Gamma\beta}
= \n{\Lambda}{\Gamma\nu} \a{\nu}{\alpha\beta} .
\label{sgcouplingrule'}
\endEQ
In this multi-graviton and multi-gravitino gauge theory,
because $\h{aBB'}{\mu}$ as well as $\rs{aB}{\Lambda}$ 
are each ordinary commuting fields, 
supersymmetry gauge invariance of the theory 
relies on the antisymmetry properties of 
the \grino/ coupling constants \eqref{sgcouplingrule},
which requires that the number $n'$ of \grino/ fields be at least two. 
(This is somewhat analogous to aspects of the structure of \YM/ gauge theory 
for a set of spin-1 fields.)
These coupling constants in the field-coupling algebra $\multA{}$,
moreover, can be generalized to become complex valued 
with antisymmetry properties that now involve complex conjugation 
\EQ
\l{\mu}{\Lambda\Gamma} = -\cl{\mu}{\Gamma\Lambda} 
\neq -\l{\mu}{\Gamma\Lambda} ,\quad
\n{\Lambda}{\Gamma\nu} \l{\nu}{\Sigma\Omega} 
= \n{\Lambda}{\Omega\nu} \cl{\nu}{\Gamma\Sigma}
\neq \n{\Lambda}{\Omega\nu} \l{\nu}{\Gamma\Sigma} .
\endEQ

The previous two types of multi-graviton and multi-gravitino theories
\eqsref{sgweylfield}{sgLagr}
can be reformulated simply in terms of 
an $\hrsA{even}$-valued tetrad field
and an $\hrsA{odd}$-valued vector-spinor field
\EQ
\algframe{a}{BB'}= \openone \invsodder{a}{BB'} + \algh{a}{BB'} ,\quad
\algweyl{a}{B}= \algrs{a}{B}
\label{algtetradweyl}
\endEQ
through employing a unit element $\unit{\mu}$ in $\hrsA{}$
(appending one if none exists \cite{Wald}). 
This tetrad $\algframe{a}{BB'}$ 
is related to the \gron/ field variable \eqref{sgefield}
in a similar way to the metric tensor field \eqrefs{gfield}{algmetric}
in the Cutler-Wald multi-graviton theories. 
An analogous relation is evident between 
the vector-spinor $\algweyl{a}{B}$
and \grino/ field variable \eqref{sgweylfield}. 

A more interesting and highly novel 
multi-graviton and multi-gravitino gauge theory is obtained
when $\multA{even}$ consists of
formal rules of anticommutative, anti-associative multiplication
while $\multA{odd}$ consists of
standard commutative, associative multiplication rules. 
This makes the \gron/ fields $\h{aBB'}{\mu}$
each formally anticommuting
and the \grino/ fields $\rs{aB}{\Lambda}$
each formally commuting,
as summarized by the rules \eqref{gnewmultrule}
together with 
\EQs
&&
\vartheta_1 \vartheta_2 = \vartheta_2 \vartheta_1 ,\quad
\vartheta_1 (\vartheta_2 \vartheta_3) 
= (\vartheta_1 \vartheta_2) \vartheta_3 ,
\label{sgmultrule}\\
&&
\vartheta_1 (\varrho_1 \varrho_2) 
= (\vartheta_1 \varrho_1) \varrho_2 ,\quad
(\vartheta_1 \vartheta_2) \varrho_1
= \vartheta_1 (\vartheta_2 \varrho_1) 
= (\vartheta_1 \varrho_1) \vartheta_2 ,
\label{sgmultrule'}
\endEQs
where $\vartheta$'s stand for $\rs{aB}{\Lambda}$
or odd-graded products of $\rs{aB}{\Lambda}$'s and $\h{aBB'}{\mu}$'s,
while $\varrho$'s stand for $\h{ab}{\mu}$ 
or even-graded products of $\rs{aB}{\Lambda}$'s and $\h{aBB'}{\mu}$'s.
These formal multiplication rules 
\eqref{gnewmultrule}, \eqrefs{sgmultrule}{sgmultrule'}
reverse the usual spin-statistics relation at the classical level,
and therefore we will refer to them as 
reverse-\Gra/ graded-associative multiplication rules. 
At the same time, 
the field-coupling algebra $\hrsA{}$ is given by 
an arbitrary $n+n'$ dimensional 
anticommutative, graded-associative algebra,
whose $n$ dimensional even part $\hrsA{even}$ is anti-associative. 
The structure constants 
$\a{\mu}{\alpha\beta}$, 
$\l{\mu}{\Omega\Gamma}$,
$\n{\Lambda}{\Gamma\alpha}$
of $\hrsA{}$ 
are characterized by the relations \eqref{gnewcouplingrule},
\eqrefs{sgcouplingrule}{sgcouplingrule'}. 
It is straightforward to show that 
the antisymmetry relations \eqrefs{gnewcouplingrule}{sgcouplingrule} imply 
\EQ
\a{\mu}{\nu\sigma} \l{\nu}{\Lambda\Gamma} \l{\sigma}{\Sigma\Omega} =0 .
\label{sgcouplingrule''}
\endEQ
For these relations to be satisfied, 
there is, in general, a tradeoff between 
the degree of nontriviality of $\l{\mu}{\Omega\Gamma}$
compared with $\n{\Lambda}{\Gamma\alpha}$ and $\a{\mu}{\alpha\beta}$. 
We consider two simple examples. 
First, 
$\l{1}{1\ 2} \neq 0$, $\l{2}{1\ 3} \neq 0$, $\l{3}{2\ 3} \neq 0$,
$\l{4}{1\ 4} \neq 0$, $\l{5}{2\ 4} \neq 0$, $\l{6}{3\ 4} \neq 0$, 
and $\n{4}{1\ 3} = - \n{4}{2\ 2} = \n{4}{3\ 1} \neq 0$, 
$\n{5}{1\ 5} = - \n{5}{2\ 4} = \n{5}{4\ 1} \neq 0$, 
$\n{6}{2\ 6} = - \n{6}{3\ 5} = \n{6}{4\ 3} \neq 0$, 
(all others zero, taking into account antisymmetry) 
with $n'=8,n=6$, 
which produces a fourteen-dimensional anticommutative, associative algebra
where the entire even subalgebra is most nilpotent. 
(This example arises by forming all possible 
anticommutative products of four odd generators:
$o_1,o_2,o_3,o_4,
o_{12}\equiv o_1 o_2, o_{13}\equiv o_1 o_3, 
o_{23}\equiv o_2 o_3, o_{14}\equiv o_1 o_4, 
o_{24}\equiv o_2 o_4, o_{34}\equiv o_3 o_4, 
o_{123}\equiv o_1 o_{23} = -o_2 o_{13} =o_3 o_{12}, 
o_{124}\equiv o_1 o_{24} = -o_2 o_{14} =o_4 o_{12}, 
o_{134}\equiv o_1 o_{34} = -o_3 o_{14} =o_4 o_{13}, 
o_{234}\equiv o_2 o_{34} = -o_3 o_{24} =o_4 o_{23}$
such that $o_{12} o_{34} = o_{13} o_{24} = o_{23} o_{14}=0$,
and all other products equal to zero.)
Second, 
$\a{3}{1\ 2} \neq 0$,
and $\n{1}{2\ 1}  \neq 0$, $\n{3}{4\ 1} \neq 0$, 
$\l{2}{2\ 4}  \neq 0$, $\l{3}{1\ 4} = \l{3}{2\ 3} \neq 0$,
(all others zero, taking into account (anti)symmetry)
with $n=3,n'=4$,
producing a seven-dimensional associative, anticommutative algebra. 
(This example arises by forming all possible 
anticommutative even-graded products of a single even generator
and four odd generators:
$u,o_1,o_2,o_3,o_4$
such that 
$u o_2= o_1, u o_3= o_4, 
o_{14}\equiv o_1 o_4, o_{23}\equiv o_2 o_3, o_{24}\equiv o_2 o_4$
and hence 
$u o_{24} = o_{14}= -u o_{42} = o_{23}$, 
with all other products equal to zero.)
It is easy to extend these two examples to obtain 
an anticommutative, graded-associative algebra
by enlarging the even subalgebra to make it anti-associative
(namely, via appending additional even generators). 

Compared with the supersymmetric extensions of 
the Cutler-Wald  multi-graviton theories, 
note that the new multi-graviton and multi-gravitino theories 
presented here 
do not have a formulation in terms of 
$\hrsA{}$-valued fields \eqref{algtetradweyl},
since no anticommutative algebra can admit a unit element. 

Finally, 
a scalar Lagrangian 
for these multi-graviton and multi-gravitino gauge theories
\eqsref{sgweylfield}{sgLagr}
is readily obtained by the same kind of procedure
as summarized in Propositions~3 to~5
for the multi-graviton gauge theories \eqsref{gfield}{gLagr}. 

\Proclaim{ Proposition~6. }{
(i) 
For $\hrsA{}$ given by 
a commutative, associative graded algebra with a unit element, 
a scalar Lagrangian for the nonlinear theories \eqsref{sgweylfield}{sgLagr}
using formal \Gra/ multiplication rules
for $\h{aBB'}{\mu}$ and $\rs{aB}{\Lambda}$ 
is obtained by 
\EQ
\lagr{}{spin2,3/2} = \dualq{\mu}{} \lagr{\mu}{spin2,3/2}
\label{sgscalarL}
\endEQ
where $\dualq{\mu}{}$ projects onto 
a most nilpotent element (or, if none, a unit element) in $\hrsA{}$. 
\vskip0pt
(ii) 
For $\hrsA{}$ given by 
an anticommutative, graded-associative algebra, 
a scalar Lagrangian 
for the nonlinear theories \eqsref{gfield}{gLagr} 
using formal reverse-\Gra/ graded-associative multiplication rules
for $\h{aBB'}{\mu}$ and $\rs{aB}{\Lambda}$ 
is constructed in terms of components 
$\q{\mu\nu}{}=\q{[\mu\nu]}{}$, 
$\othq{\Lambda\Gamma}{}= \othq{[\Lambda\Gamma]}{}$
of an invariant nondegenerate inner product on $\hrsA{}$
by 
\EQ
\lagr{}{spin2,3/2} 
= \q{\mu\nu}{} \lagr{\mu\nu}{spin2}
+ \othq{\Lambda\Gamma}{} \lagr{\Lambda\Gamma}{spin3/2}
\label{sgnilpscalarL}
\endEQ
where $\lagr{\mu\nu}{spin2}$ and $\lagr{\Lambda\Gamma}{spin3/2}$
are the coefficients defined from 
$\lagr{\sigma}{spin2,3/2} 
= \a{\sigma}{\mu\nu} \lagr{\mu\nu}{spin2} 
+ \l{\sigma}{\Lambda\Gamma} \lagr{\Lambda\Gamma}{spin3/2} +$
linear $\h{aBB'}{\sigma}$ terms. 
\vskip0pt
(iii) 
For $\hrsA{}$ given by 
a graded-commutative, associative algebra, 
a scalar Lagrangian 
for the nonlinear theories \eqsref{gfield}{gLagr} 
using ordinary multiplication rules
for $\h{aBB'}{\mu}$ and $\rs{aB}{\Lambda}$ 
is obtained from either \eqref{sgscalarL}
if $\hrsA{}$ possesses a unit element, 
or \eqref{sgnilpscalarL} if $\hrsA{}$ is completely nilpotent. 
}

In case $(i)$, 
the most nilpotent ideal in $\hrsA{even}$ is required to be one-dimensional, 
which holds if $\hrsA{}$ is any even-\Gra/ graded algebra
such that the number of generators is an even integer. 
In case $(ii)$, 
we note that the invariance property of the inner product 
is characterized by the relation \eqref{gskewmetricrule} 
together with 
\EQ
\q{\nu\mu}{} \l{\mu}{\Lambda\Gamma} 
=  \othq{\Lambda\Omega}{} \n{\Omega}{\Gamma\nu} .
\endEQ
Further details will not be pursued for this case other than to remark that,
as in the situation for anticommutative, anti-associative algebras $\hrsA{}$
discussed in \secref{gravitonth}, 
invariance and nondegeneracy of an inner product 
necessarily forces some restrictions on 
the most nilpotent ideals associated with even and odd multiplication 
in $\hrsA{}$. 
Case (iii) obviously requires the same conditions on $\hrsA{}$ 
as noted in cases (i) and (ii).

\section{ Concluding remarks }
\label{remarks}

This paper has obtained a complete classification of
the possibilities allowed for 
non-higher-derivative classical gauge theories of
a nonlinearly coupled set of any number of \gron/ fields and \grino/ fields 
with general formal internal multiplication rules.
The classification includes, as a special case, 
all allowed possibilities for 
a nonlinearly coupled set of any number of \gron/ fields alone. 
This classification in the \gron/ case has found a novel type of
nonlinear gauge theory for two or more \gron/ fields,
with coupling constants given by 
anticommutative, anti-associative algebras,
and using formal multiplication rules 
that make the \gron/ fields anticommuting 
(while products obey anti-associativity). 
Moreover, apart from the previously known type of nonlinear gauge theories
found by Cutler and Wald 
for one or more ordinary commuting \gron/ fields
whose coupling constants arise from commutative, associative algebras,
the classification shows that there are no other possibilities. 
Supersymmetric extensions of these nonlinear theories have been obtained
and further proved to be the only allowed possibilities 
in the more general case when \grino/ fields are included. 

The framework used for obtaining these main results
consists of a deformation analysis of the linear gauge theory
for a set of uncoupled \gron/ and \grino/ fields
with a common internal vector space structure for each field
as necessary for accommodating formal multiplication rules. 
Determining equations for deformation terms have been formulated and solved 
by the methods of \Ref{AMSpaper,Annalspaper}
generalized to the case of more than one \gron/ field 
and more than one \grino/ field. 
It is important to emphasize that
no assumptions or special conditions were imposed here
on the form considered for possible deformation terms
except for requiring these terms to involve 
no higher derivatives than those appearing in the linear field equations
and abelian gauge symmetries. 

The results and framework here 
give a culmination of much previous work in the literature
on non-higher-derivative deformations for gauge theories of \gron/ fields
\cite{uniqueness1,uniqueness2,uniqueness3,uniqueness4,uniqueness5,
uniqueness6,uniqueness7,Cutler-Wald,Annalspaper,Henneaux2},
including couplings to \grino/ fields
\cite{sguniqueness1,sguniqueness2,sguniqueness3,sg3},
in four spacetime dimensions. 
In particular, 
uniqueness results for such deformations 
are strengthened in this paper
through the use of a nonsymmetric tensor analogous to a tetrad
for the \gron/ field variable,
which has given a strong no-go theorem on possibilities
for deforming the local Lorentz symmetry on the \gron/ field variable
(and \grino/ field variable), 
and for producing couplings that 
involve the skew part of the \gron/ field variable. 
At the same time, 
this paper considerably generalizes all previous frameworks 
by encompassing the most general possibilities for formal rules
other than commutative, associative multiplication of \gron/ fields,
and its extension to \Gra/ multiplication of \grino/ fields, 
which has uncovered the novel existence of consistent nonlinear couplings
for more than one anticommuting \gron/ field
and supersymmetric extensions with more than one commuting \grino/ field. 

Some further lines of work are suggested by the main results discussed above. 
Do the new multi-graviton and multi-gravitino gauge theories
found here have a well-posed initial value formulation
as classical field theories?
This could be expected to hold by comparison with 
the initial value formulation of \NSG/ as a single graviton and gravitino 
classical field theory. 
Is there a consistent quantization for the new gauge theories
with the multi-graviton and multi-gravitino quantum fields 
obeying reversed spin-statistics commutation relations
that mirror the formal multiplication rules found here at the classical level?
This could be investigated first for a model nonlinear theory of,
for example, coupled anticommuting scalar fields 
and commuting neutrino or Dirac fields
as obtained in a manner analogous to way in which 
the new  multi graviton and gravitino gauge theories
are derived from the formulation of \NSG/ for an algebra-valued 
\gron/ field and \grino/ field.

\appendix*
\section{ Graded algebras }

In this appendix, 
some definitions and structure of graded algebras are summarized. 
Throughout let $\genA{}$ be an algebra possessing a grading that determines
a decomposition into even and odd subspaces 
$\genA{}=\genA{even}\oplus \genA{odd}$
such that products of two even elements are even, 
products of two odd elements are even, 
and products of one even element and one odd element are odd. 

A unit element $\openone$, if one exists, is an element in $\genA{even}$
such that $\openone x= x\openone =x$ for all elements $x$ in $\genA{}$.
A non-unit element element $x$ in $\genA{}$ is nilpotent 
if the $p$-fold product of itself, denoted $x^p$, 
vanishes for some integer $p>1$. 
A most nilpotent element $\genA{}$ is a nilpotent element $y$ 
whose product with all other nilpotent elements $x$ vanishes, $xy=0$. 
$\genA{}$ is called completely nilpotent if every element in $\genA{}$
is nilpotent. 

$\genA{}$ is an associative algebra 
if all multiplication obeys associativity,
namely, $x(yz)= (xy)z$ for all elements $x,y,z$ in $\genA{}$.
An associative algebra $\genA{}$ is (anti) commutative if 
all multiplication is (anti) commutative, 
namely, $xy= (-)yx$ for all elements $x,y$ in $\genA{}$.
An associative algebra $\genA{}$ is graded-commutative if 
odd multiplication is anticommutative,  
even multiplication is commutative, 
and even-odd multiplication is commutative, 
namely, 
$xy=yx$ for all elements $x,y$ in $\genA{even}$, 
$xy=-yx$ for all elements $x,y$ in $\genA{odd}$, 
and $xy=yx$ for all elements $x$ in $\genA{even}$ and $y$ in $\genA{odd}$. 

A nilpotent \Gra/ algebra $\gradedG{nilpotent}$ 
is an associative, graded-commutative algebra $\genA{}$ 
formed by the span of all possible products of some number of odd generators,
namely, 
$o_1,o_2,\ldots$, 
$o_{12}\equiv o_1 o_2, o_{13}\equiv o_1 o_3, o_{23}\equiv o_2 o_3,\ldots$,
$o_{123}\equiv o_1 o_2 o_3,\ldots$, etc.
A \Gra/ algebra $\gradedG{unit}$ 
is the enlargement of a nilpotent \Gra/ algebra
obtained by appending a unit element $\openone$. 
A (nilpotent) even-\Gra/ algebra 
is an associative, commutative algebra isomorphic to 
the even subalgebra $\gradedG{even}$ of a (nilpotent) \Gra/ algebra. 
Equivalently, a (non-nilpotent) even-\Gra/ algebra is 
formed by the span of (a unit element together with) 
all possible products of some number of even generators whose square is zero,
namely, 
$u_1,u_2,u_3,\ldots$,
$u_{12}\equiv u_1 u_2, u_{13}\equiv u_1 u_3, u_{23}\equiv u_2 u_3,\ldots$,
$u_{123}\equiv u_1 u_2 u_3,\ldots$, etc.

$\genA{}$ is an anticommutative, graded-associative algebra 
if products of two elements are anticommutative, 
and products of three or more even elements are anti-associative, 
while all other products are associative, 
namely, $xy=-yx$, 
$x(yz)= -(xy)z$ for all elements $x,y,z$ in $\genA{even}$,
and $x(yz)= (xy)z$ for all elements where at least one of $x,y,z$ 
is in $\genA{odd}$. 

A reverse-\Gra/ graded-associative algebra is defined here to be 
an algebra $\genA{}$ such that 
even multiplication is anticommutative and anti-associative, 
odd multiplication and even-odd multiplication are each 
commutative and associative. 

The set of most nilpotent elements in an algebra $\genA{}$ forms an ideal,
while the set of nilpotent elements also forms an ideal 
if multiplication is either (anti) associative or graded-associative, 
and if multiplication is either (anti) commutative or graded-commutative.

\end{document}